\documentclass[10pt,journal,twocolumn]{IEEEtran}
\ifCLASSINFOpdf
   \usepackage[pdftex]{graphicx}
   \DeclareGraphicsExtensions{.pdf,.jpeg,.png}
\else
   \usepackage[dvips]{graphicx}
   \DeclareGraphicsExtensions{.eps,.pdf}
\fi
\usepackage{amsmath}
\usepackage{amssymb}
\usepackage{amsthm}
\usepackage{cancel}
\usepackage{breqn}
\usepackage{multirow}
\usepackage{rotating}
\usepackage{verbatim}
\usepackage{textcomp,gensymb}
\usepackage[backend=biber,style=ieee,doi=false,isbn=false]{biblatex}
\ifCLASSOPTIONcompsoc
    \usepackage[caption=false, font=normalsize, labelfont=sf, textfont=sf]{subfig}
\else
\usepackage[caption=false, font=footnotesize]{subfig}
\fi

\DeclareMathOperator*{\argmin}{arg\,min}

\newcommand{\Expect}{{\rm I\kern-.5em E}}
\newcommand{\multiline}[2][c]{\begin{tabular}[#1]{@{}c@{}}#2\end{tabular}}
\newcommand*\dif{\mathop{}\!\mathrm{d}}
\DeclareMathOperator*{\biglor}{\lor}
\DeclareMathAlphabet{\mathitsf}{\encodingdefault}{\sfdefault}{m}{sl}
\DeclareMathAlphabet{\mathitbfsf}{\encodingdefault}{\sfdefault}{bx}{sl}
\usepackage{cool}
\begin{document}

\title{3D Polarized Modulation: System Analysis and Performance}

\author{Pol Henarejos,%
        ~Ana I. P\'erez-Neira
\thanks{Pol Henarejos is with the Centre Tecnol\`ogic de Telecomunicacions de Catalunya.}
\thanks{Ana I. P\'erez-Neira is with the Centre Tecnol\`ogic de Telecomunicacions de Catalunya and the Universitat Polit\`ecnica de Catalunya.}%
}

\maketitle

\begin{abstract}
In this paper we present a novel modulation technique for dual polarization communication systems, which reduces the error rate compared with the existent schemes. This modulation places the symbols in a 3D constellation, rather than the classic approach of 2D. Adjusting the phase of these symbols depending on the information bits, we are able to increase the bit rate. Hence, the proposed scheme conveys information by selecting both polarization state and the phase of radiated electromagnetic wave. We also analyse the performance of 3D Polarized Modulation (PMod) for different constellation sizes and we obtain a curve of rate adaptation. Finally, we compare the proposed 3D PMod with other existing schemes such as single polarization Phase Shift Keying (PSK) and double polarization Vertical Bell Laboratories Layer Space-Time (V-BLAST), both carrying the same number of information bits. The results show that 3D PMod always outperforms all other schemes, except for low order modulation. Therefore, we can conclude that 3D PMod is an excellent candidate for medium and high modulation order transmissions.
\end{abstract}

\begin{IEEEkeywords}
Polarized Modulation,3D Modulation
\end{IEEEkeywords}

\IEEEpeerreviewmaketitle

\section{Introduction}
Polarization dimension is being used massively in satellite communications during many decades. Until few years, transmissions took place using fixed and orthogonal polarization schemes, vertical/horizontal or left-hand circular polarization (LHCP) / right-hand circular polarization (RHCP). However, with the explosion of digital signal processing, the dynamic use of polarization schemes is taking an important role to increase the throughput and robustness. 

Standards, such as Broadband Global Area Network (BGAN) or Digital Video Broadcast S2 (DVB-S2), are including the use of multiple polarization schemes in the incoming releases because the benefits of using dynamic polarization has been discussed in many works \cite{Arapoglou2010,Arapoglou2011}. Schemes, such as Orthogonal Space-Time Block Coding (OSTBC) or Vertical Bell Laboratories Layer Space-Time (V-BLAST), are studied replacing the spatial domain with polarization domain. In the major part of use cases, these schemes produce higher throughputs when they are compared with single polarization cases \cite{Henarejos2013}. Moreover, the benefits of dynamic reconfiguration of dual polarization schemes are also proven in \cite{Henarejos2016}.

Polarization Shift Keying (PolSK) is also a double polarization scheme, which conveys information not by radiating a symbol modulated with in-phase and quadrature (I/Q) model, but in the selected polarization. In this scheme, the information bits determine which polarization is used to transmit a tone \cite{Benedetto1992,Arend2016}. 

Among the aforementioned schemes, Polarized Modulation (PMod) \cite{Henarejos2015a} conveys the information not only in the polarization hop, as with PolSK, but also radiating an electromagnetic wave using in-phase/quadrature symbols, which are chosen from a modulation set depending on the information. This scheme has also the equivalent with the spatial domain, the so-called Spatial Modulation \cite{Renzo2014}. However, though in the spatial domain an arbitrary number of uncorrelated antennas can be used, in the polarization domain only two uncorrelated polarizations are available. This constraint penalizes the throughput and is specially indicated for low order modulation transmissions \cite{Henarejos2017}. Authors of \cite{Guo2017} present an exhaustive survey of all known technologies based on polarization. Besides PMod, there are other solutions that cover higher spectral efficiencies intended for high Signal to Noise Ratio (SNR) regimes.

In the present work, we introduce a PMod scheme using an arbitrary number of polarizations, which we entitled 3D Polarized Modulation (3D PMod) targeted to satellite networks. We emphasize that the term \emph{3D} is related with the dimension of the space where the constellation lays. Whereas this space is two-dimensional in classical modulations, in our work presented in this paper this space is three-dimensional. Note that this concept is not analogous to the spatial case, where the \emph{3D} term refers to the spatial coverage in azimuth and elevation \cite{Elkawafi2017,Fu2016}. In the polarization case, by using a 3D sphere as the constellation, we are able to map 3D points with the respective electric field. Hence, compared with the classic 2D I/Q constellation mapping, placing the symbol on a sphere increases the minimum distance between symbols. Therefore, we can reduce the error rate and increase the throughput without requiring additional energy.

Among the aforementioned schemes that use dual polarization, the authors in \cite{Gu2017} employ the quaternary modulation by using two orthogonal polarizations and their respective cross-polarizations. In \cite{Cao2011} this modulation is used to enable multiple access to serve different number of users. In \cite{Zafari2017}, a hybrid solution between spatial and polarized modulation is described for two double polarized antennas. In \cite{Guo2016}, although the polarization dimension is not provided, the Spatial Modulation is described in terms of 3-D mapping. Furthermore, the work of \cite{Zhang2017a} extends the contribution of \cite{Benedetto1992} and a computational complexity analysis is provided. In \cite{Nie2014}, the authors employ the polarization dimension for common phase error in Orthogonal Frequency Division Multiplexing systems.

Other works, such as \cite{Wei2012,Wei2013,Wei2013a,Wei2016}, introduce the concept of 3D Polarized Modulation based on sphere mapping for pre-compensating channel impairments. In contrast to these works, we denote the following novel contribution:
\begin{enumerate}
	\item We describe the transmitter signal processing chain and we propose two different receivers, based on computational complexity. 
	\item Whereas these works use a non-optimal mapping, we employ a mapping on the sphere that guarantees the maximum minimum distance in order to minimize the error rate.
	\item We introduce the upper bound of Bit Error Rate (BER), which derives into a closed-form expression and becomes a tight bound, as we prove in the Results section. Additionally, this expression more compact compared with the instantaneous expression provided by \cite{Wei2012,Wei2013a}.
	\item The sphere modulation that we employ requires normalized energy. Since Quadrature Amplitude Modulation (QAM), used by \cite{Wei2012,Wei2013,Wei2013a,Wei2016}, is a non unit energy scheme, we use modulations in phase, such as PSK, to preserve this constraint.
	\item We compare and benchmark our proposals with other solutions such as Lattice Amplitude Modulation (LAM), Polarization Multiplexing or Single Polarization.
\end{enumerate}

\section{Sphere Modulation}
Prior to the description of the proposed modulation, we introduce the concept of sphere modulation. Traditionally, the mapping between bits of information and electromagnetic propagation is performed by using a two-dimensional plane, where x-y axis represent the in-phase and quadrature (I/Q) components in the baseband model. In the case of mapping between bits and polarization state, the x-y axis represent two orthogonal polarizations, such as V/H or RHCP/LHCP. In this section, we extend to three dimensional case the mapping between bits and polarization state by using a sphere. In this section, we describe this mapping needed to enable the Polarized Modulation.

In $1852$, Sir George Gabriel Stokes discovered that the polarization state of any electromagnetic wave can be characterized by four parameters, now called Stokes parameters \cite{Stokes1852}. These parameters can be represented with the Stokes vector. Stokes parameters are obtained by decomposing the $\mathbf{E}$ into two orthogonal components. Given an electromagnetic wave propagating along z-axis, the electric field $\mathbf{E}$ can be uniquely decomposed into two orthogonal components $E_x$ and $E_y$ with $\mathbf{x}/\mathbf{y}$ as the reference basis. Thus, if the angular frequency is denoted by $\omega$, it can be expressed as
\begin{equation}
\begin{split}
\mathbf{E}(z,t)&=\Re\left\{E_x(z,t)\mathbf{x}+E_y(z,t)\mathbf{y}\right\}\\
E_x(z,t)&=E_{0x}e^{j\left(\omega t-kz+\varphi_x\right)}=E_xe^{j\omega t-kz}\\
E_y(z,t)&=E_{0y}e^{j\left(\omega t-kz+\varphi_y\right)}=E_ye^{j\omega t-kz}
\end{split}
\end{equation}
where $E_{0x}$ and $E_{0y}$ are the amplitude of each component, $k$ is the wavenumber and $\varphi_x$ and $\varphi_y$ are their respective phases. The contribution of $\mathbf{E}_0$ can be decomposed by $E_{0x}$ and $E_{0y}$ as follows
\begin{equation}
\mathbf{E}_0=\begin{pmatrix}E_x\\E_y\end{pmatrix}=\begin{pmatrix}E_{0x}e^{j\varphi_x}\\E_{0y}e^{j\varphi_y}\end{pmatrix}.
\end{equation}
The vector $\mathbf{E}_0$ is the Jones vector.

The four Stokes parameters describe the state of polarization and are defined as
\begin{equation}
\begin{split}
S_0&=\left|E_x\right|^2+\left|E_y\right|^2=E_{0x}^2+E_{0y}^2\\
S_1&=\left|E_x\right|^2-\left|E_y\right|^2=E_{0x}^2-E_{0y}^2\\
S_2&=E_xE_y^*+E_x^*E_y=2E_{0x}E_{0y}\cos\theta\\
S_3&=j\left(E_xE_y^*-E_x^*E_y\right)=2E_{0x}E_{0y}\sin\theta
\end{split}
\end{equation}
where $\theta=\varphi_x-\varphi_y$. Hence, the Stokes vector, denoted as $\mathbf{S}$, is represented as
\begin{equation}
\mathbf{S}=\begin{pmatrix}S_0\\S_1\\S_2\\S_3\end{pmatrix}.
\end{equation}
Additional information is provided in \cite{Guo2017}.
Stokes components are real quantities and not only characterize the state of the polarization but also indicate the degree of polarization with the following inequality
\begin{equation}
S_0^2\geq S_1^2+S_2^2+S_3^2,
\label{eq:polineq}
\end{equation}
which is fulfilled with equality when the electromagnetic wave is completely polarized. 

$S_1$, $S_2$ and $S_3$ span a three-dimensional polarization space and therefore all polarization states can be expressed as a linear combination of these parameters. Note that, even in the case where the wave is partially polarized, the wave can be decoupled into a contribution of a completely polarized wave and an unpolarized light as follows
\begin{equation}
\mathbf{S}=(1-P)\begin{pmatrix}S_0\\0\\0\\0\end{pmatrix}+P\begin{pmatrix}S_0\\S_1\\S_2\\S_3\end{pmatrix},
\end{equation}
where $P=\frac{\sqrt{S_1^2+S_2^2+S_3^2}}{S_0}$, $0\leq P\leq 1$, is the degree of polarization. The practical consequence of partially polarized waveform is that the power of completely polarized waveform is reduced by a factor of $P$. Since the unpolarized contribution does not affect to the proposed scheme, hereinafter we focus on the completely polarized waveform, whose SNR is affected by $P$. Note that, although the signal is partially polarized, it is still affected by the noise, even if it is unpolarized.

\begin{figure}[!ht]
\centering
\subfloat[$L=2$]{\includegraphics[width=0.25\linewidth,clip=true,bb=90 110 290 330]{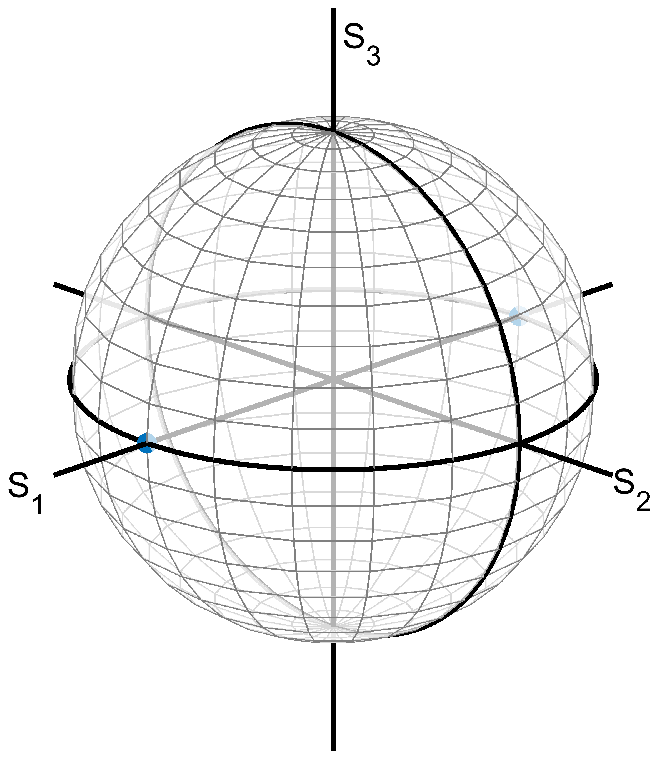}\label{fig:sphere2}}\hfill
\subfloat[$L=4$]{\includegraphics[width=0.25\linewidth,clip=true,bb=90 110 290 330]{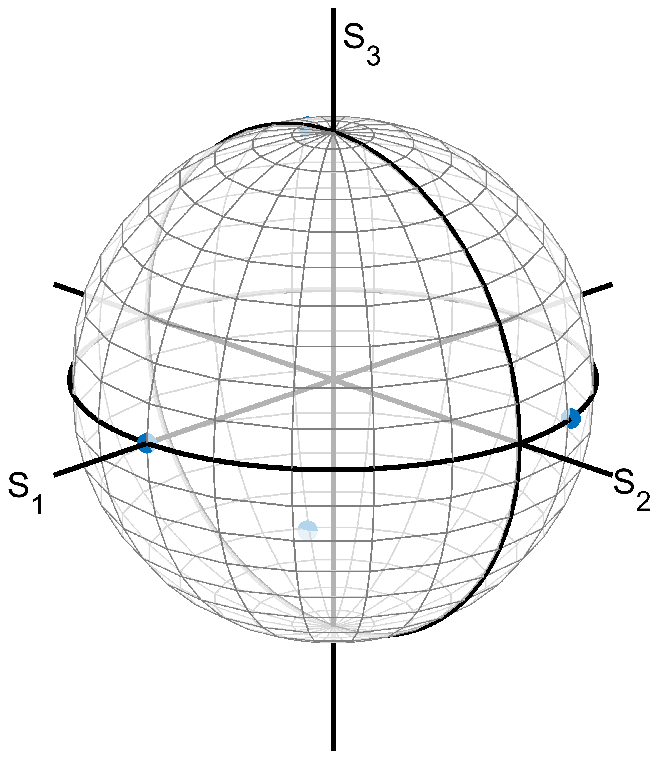}\label{fig:sphere4}}\hfill
\subfloat[$L=8$]{\includegraphics[width=0.25\linewidth,clip=true,bb=90 110 290 330]{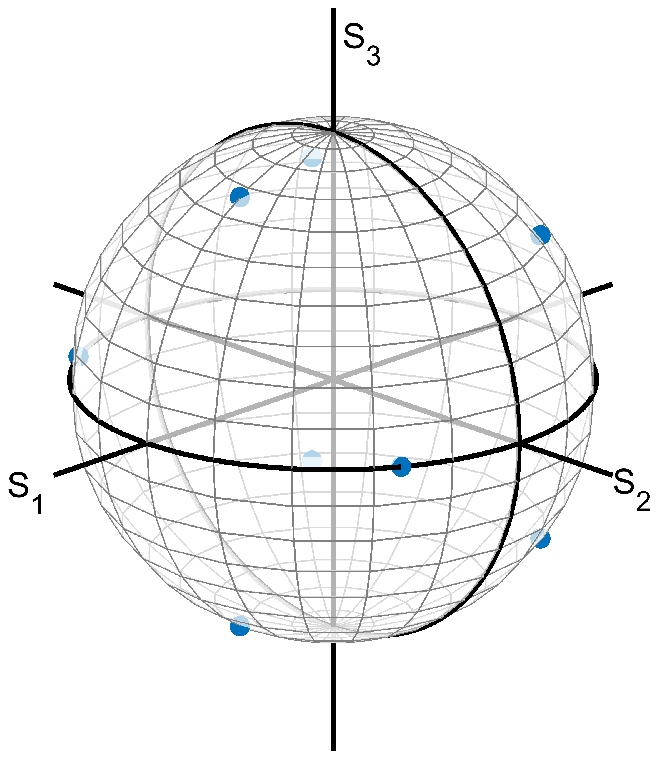}\label{fig:sphere8}}\hfill
\subfloat[$L=16$]{\includegraphics[width=0.25\linewidth,clip=true,bb=90 110 290 330]{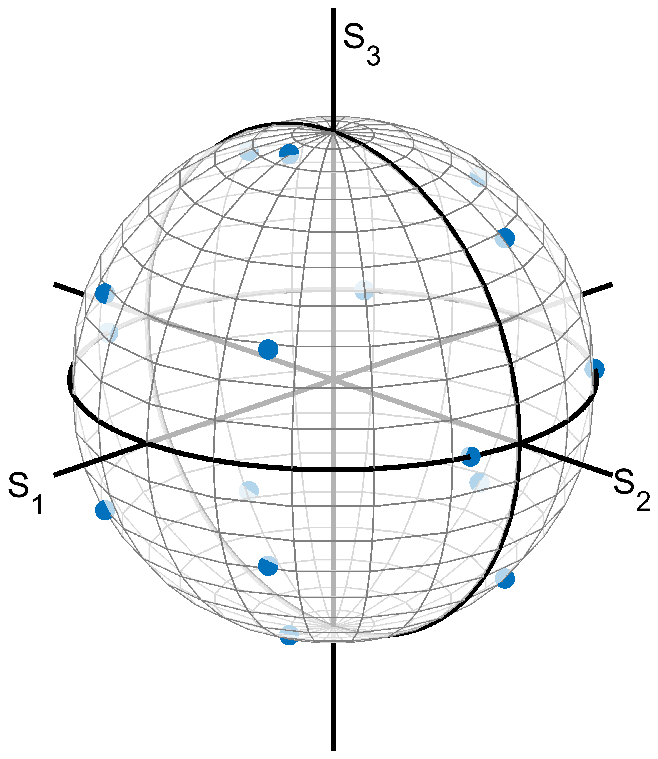}\label{fig:sphere16}}\hfill
\caption{Poincar\'e sphere facing $L$ points in such a way that the minimum distance is maximized.}
\label{fig:expoincare}
\end{figure}

Hence, the equality in~\eqref{eq:polineq}, it describes an sphere of radius $S_0$ and centred at the origin. This sphere was proposed by Henri Poincar\'e in $1892$ and now it is called the Poincar\'e Sphere \cite{Poincare1892}. For example, Fig.~\ref{fig:expoincare} displays the Poincar\'e Sphere facing $L$ points where the minimum distance is maximized, using the Sloane 3D packs~\cite{Sloane}. These points describe the polarization state used to radiate the I/Q symbol. It is important to remark that the points proposed in the present manuscript are the corrected version from \cite{Sloane}, produced using exact numbers despite of floating point comma precision. 

Expressing the Jones vector from the Stokes vector is not straightforward. The Stokes vector measures the intensities of the polarized wave, whilst the Jones vector contains information about the complex components, including the magnitude and the phase.
Using the Stokes vector we can find the following relationships
\begin{equation}
\begin{split}
S_0+S_1&=2|E_x|^2\\
S_0-S_1&=2|E_y|^2\\
\frac{S_3}{S_2}&=\tan\theta,\ \cos\theta\neq0.
\end{split}
\end{equation}
Furthermore, it is important to remark that, whereas Stokes vector is used to perform the mapping between the bits of information and the polarization state, the Jones vector is the complex electric field that is used by the orthogonal dipoles.

Hence, after some mathematical manipulations, we are able to express the Jones vector as a function of Stokes parameters by
\begin{equation}
\mathbf{E}_0=
\begin{pmatrix}\sqrt{\frac{S_0+S_1}{2}}e^{-j\theta}\\\sqrt{\frac{S_0-S_1}{2}}e^{j\theta}\end{pmatrix},
\label{eq:stokes2E}
\end{equation}
where $\theta=\frac{1}{2}\arctan{\left(\frac{S_3}{S_2}\right)}$. Note that $\arctan{\frac{y}{x}}$ is the poorer form of the argument since it is not well defined if $x=0$ and does not preserve the signs of $x$ and $y$. In order to solve it, we use the $\textrm{arctan2}(y,x)$ function instead, which is defined as
\begin{equation}
\textrm{arctan2}(y,x)=\left\{\begin{array}{ll}
\arctan\left(\frac{y}{x}\right) & \textrm{if } x>0\\
\arctan\left(\frac{y}{x}\right)+\pi & \textrm{if } x<0\textrm{ and }y\geq0\\
\arctan\left(\frac{y}{x}\right)-\pi & \textrm{if } x<0\textrm{ and }y<0\\
\frac{\pi}{2} & \textrm{if }x=0\textrm{ and }y>0\\
-\frac{\pi}{2} & \textrm{if }x=0\textrm{ and }y<0\\
0 & \textrm{if }x=0\textrm{ and }y=0.\\
\end{array}\right.
\end{equation}
and is well defined in $\mathbb{R}^2$.

The form expressed in \eqref{eq:stokes2E} is particularly interesting from the mathematical point of view when it is expressed in spherical coordinates, which takes the form
\begin{equation}
\begin{split}
S_0&=\mathcal{E}\\
S_1&=\mathcal{E}\cos\vartheta\\
S_2&=\mathcal{E}\sin\vartheta\cos\phi\\
S_3&=\mathcal{E}\sin\vartheta\sin\phi,
\label{eq:stokessph}
\end{split}
\end{equation}
where $\phi\in[0,2\pi)$ and $\vartheta\in[0,\pi)$ are the azimuthal and elevation components, respectively.

In the previous representation, by convenience, we place the $S_1$ in the z-axis, and $S_2$ and $S_3$ in the x-axis and y-axis, respectively. Moreover, we consider a sphere with its radius equal to the total energy of the symbol, $\mathcal{E}$. Thus, the Jones vector in spherical coordinates is described as
\begin{equation}
\mathbf{E}_0=
\begin{pmatrix}\sqrt{\mathcal{E}}\cos\frac{\vartheta}{2}e^{-j\frac{\phi}{2}}\\\sqrt{\mathcal{E}}\sin\frac{\vartheta}{2} e^{j\frac{\phi}{2}}\end{pmatrix}.
\label{eq:stokes2Esph}
\end{equation}

Classic digital modulations map $L$ symbols of a finite alphabet to the complex two-dimensional plane I/Q in such a way that the minimum distance is maximized to reduce the bit error rate. Using the Poincar\'e sphere, we are able to extend the same concept to the three-dimensional space. 

However, placing $L$ points on the surface of a sphere is not a straightforward problem. This problem was proposed by Tammes in $1930$ \cite{Tammes1930}, unfortunately without a closed solution. Also known as \emph{sphere packing}, this problem is also addressed in many works \cite{Hardin1995,Sloane1998,Conway2013}. Although works such as \cite{Sloane} published particular solutions that maximize the minimum distance for many dimensions and different $L$ values, there is still no closed-form expression for an arbitrary $L$ or dimension. In particular, Sloane provides solutions for $L\in[4\ldots 130]$, i.e., modulation orders from $2$ to $7$ bits.

Examining solutions provided by \cite{Sloane}, it is interesting to remark that the solutions of $L$ points where $L$ is a power of two, i.e., $L=2^{L_b}$, are simpler. In these particular cases, the points are placed by slicing the sphere and placing the points in these slices. For instance, for $L=16$, there are four slices with four points each. Due to this symmetry, it is possible to apply Gray mapping to maximize the Hamming distance.

Hence, it is possible to transmit information depending on which point on the sphere is used or, in other words, which polarization state is used. This is a more general version of the well known Polarization Shift Keying (PolSK) \cite{Benedetto1992}. 

\section{Enabling Polarized Modulation}
In this section we enable 3D Polarized Modulation by combining the mapping described in the previous section and the modulation in phase. Polarized Modulation \cite{Henarejos2015a} combines PolSK with modulated information in the amplitude and phase of the radiated waveform. This concept can also be applied using the described 3D modulation by exploiting the ambiguity of the initial phase. Our proposal focuses on choosing the initial phase by mapping certain number of bits to a PSK constellation. Hence, we are able to use two sources for transmission: the state of polarization and the initial phase. In contrast to \cite{Benedetto1992,Benedetto1994,Arend2016}, where the information is carried by the polarization shifts, we transmit information not only in the polarization shifts but also in the initial phase.

The approach that we present condenses sparse ideas from the works \cite{Wei2012,Wei2013,Wei2013a,Wei2016} under the same framework and we increase the scope of the research by introducing the optimal receiver as well as a suboptimal receiver, which is more feasible for implementation. Whereas these works are focused on the transmission pattern, we also study and analyse the design of the constellation in order to guarantee the minimum possible error rates. Furthermore, we perform extensive simulations in terms of these error rates by benchmarking with other existing solutions. 

By packing $L_b$ bits on the sphere surface and $N_b$ bits on the PSK phase we are able to convey $L_b+N_b$ bits in total. Hence, if $2^{L_b}=L$ symbols lay on the sphere and $2^{N_b}=N$ symbols are with the PSK constellation, for a particular time instant $t$, the transmitted vector as a function of Stokes parameters is described as follows
\begin{equation}
\begin{split}
\mathbf{x}[t]&=\mathbf{E}_0[t]e^{j\xi[t]}\\
&=\begin{pmatrix}\sqrt{\frac{S_0[t]+S_1[t]}{2}}e^{-j\theta[t]}\\
\sqrt{\frac{S_0[t]-S_1[t]}{2}}e^{j\theta[t]}\end{pmatrix}e^{j\xi[t]}\\
&\equiv\begin{pmatrix}E_H\\E_V\end{pmatrix},
\label{eq:Exfield}
\end{split}
\end{equation}
where $\mathbf{E}_0$ is the Jones vector of the 3D PolSK contribution. Note that the Jones vector of 3D PMod is $\mathbf{x}$ and is equal to $\mathbf{E}_0e^{j\xi}$. The transmitted vector can also be expressed in spherical coordinates as
\begin{equation}
\mathbf{x}[t]=\sqrt{\mathcal{E}}\begin{pmatrix}\cos\frac{\vartheta[t]}{2}e^{-j\frac{\phi[t]}{2}}\\
\sin\frac{\vartheta[t]}{2}e^{j\frac{\phi[t]}{2}}\end{pmatrix}e^{j\xi[t]}\equiv\begin{pmatrix}E_H\\E_V\end{pmatrix},
\label{eq:Exfieldsph}
\end{equation}
where $t$ is the time sample, $\xi[t]=\frac{2\pi}{N}n[t]$ is the modulated initial phase, $n[t]$ is the symbol with the PSK constellation modulated in phase, $S_0[t]=\sqrt{S_1[t]^2+S_2[t]^2+S_3[t]^2}=\mathcal{E}$ is the total energy transmitted by the symbol, $\theta[t]=\frac{1}{2}\arctan\left(\frac{S_3[t]}{S_2[t]}\right)$ and $S_1[t]$, $S_2[t]$, $S_3[t]$ are the coordinates of the point on the sphere surface. The expression in spherical coordinates is particularly interesting because it can be seen that $\mathbf{E}_0^H\mathbf{E}_0=1,\ \forall \phi,\vartheta$.

It is worth mentioning that introducing the $e^{j\xi[t]}$ component does not affect the computation of Stokes parameters, since it is an invariant transformation and, thus, is independent from the PolSK modulation.

\begin{figure}[!ht]
\centering
\includegraphics[width=0.9\linewidth,clip=true]{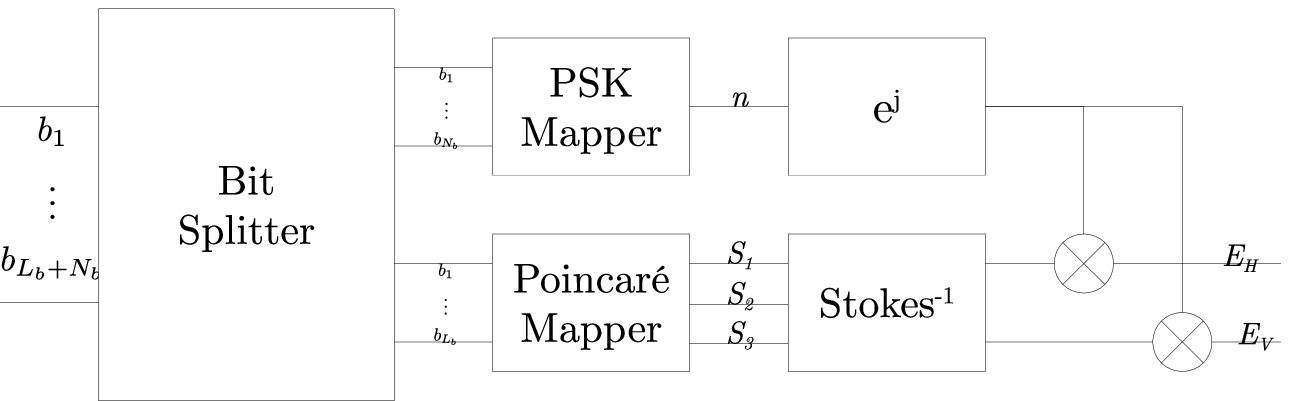}
\caption{Block diagram of 3D Polarized Modulation transmitter.}
\label{fig:3DPMod_diagram}
\end{figure}

Hence, we can describe the system model as
\begin{equation}
\mathbf{y}[t]=\mathbf{H}[t]\mathbf{x}[t]+\mathbf{w}[t],
\end{equation}
where $\mathbf{y}\in\mathbb{C}^2$ is the received signal, $\mathbf{H}\in\mathbb{C}^{2\times 2}$ is the channel matrix and $\mathbf{w}\in\mathbb{C}^2$ is the zero-mean noise vector with a covariance $\mathbf{R_w}=N_0\mathbf{I}$. Fig. \ref{fig:3DPMod_diagram} illustrates the block diagram of 3D Polarized Modulation transmitter, where $E_V$ and $E_H$ denote the vertical and horizontal electric field, respectively, and where $E_H\equiv E_x$ and $E_V\equiv E_y$ for simplicity.

One important advantage of using PSK modulation for the initial phase is that it does not affect the demodulation of Stokes parameters, since they are only affected by the differential phase. Thus, we can decode the grouped bits independently from each other. 

Based on that, we introduce two different classes of receivers:
\begin{enumerate}
	\item Joint receiver: it decodes the symbols from the PSK constellation and the Stokes parameters jointly. 
	\item Cascade receiver: it is composed by two independent receivers, faced in cascade. First, the Stokes receiver computes the Stokes parameters and second, these are used by the PSK receiver.
\end{enumerate}

\subsection{Joint Receiver}
This receiver decodes all symbols and bits without decoupling the PSK and 3D PolSK contributions. The optimal receiver implements the maximum likelihood (ML) algorithm. 

The expression of this receiver is denoted by
\begin{equation}
\begin{split}
\left(\hat{l},\hat{n}\right)&=\argmin_{l,n}\|\mathbf{y}[t]-\mathbf{H}[t]\mathbf{x}[t]\|\\
&=\argmin_{l,n}\mathbf{y}[t]^H\mathbf{x}_{l,n},
\end{split}
\end{equation}
where $\|\cdot\|$ is the $\ell^2$-norm, $\mathbf{x}_{l,n}$ is the symbol $\mathbf{x}$ using the $l$th symbol of the Poincar\'e sphere and the $n$th symbol of the PSK constellation. 

Note that, in this receiver, the search space is $L\times N$ and, hence, its computational complexity is $o(L\times N)$. 

\subsection{Cascade Receiver}
In order to reduce the complexity next we propose a suboptimal receiver. This receiver decouples the signal into two contributions: the PSK and 3D PolSK. Each contribution is decoded by an independent receiver. The PSK contribution does not affect the 3D PolSK, since the Stokes parameters are obtained using the difference of the phases of each component. Thus, the Stokes parameters can be estimated using the received signal $\mathbf{y}[t]$ straightforwardly. However, the contribution of 3D PolSK affects the PSK contribution. To estimate the phase $\hat{\xi}[t]$ we filter the received signal by a linear filter 
\begin{equation}
\hat{\xi}[t]=\arg\hat{r}[t]=\arg\left(\mathbf{a}^H[t]\mathbf{y}[t]\right),
\end{equation}
where $\mathbf{a}$ is the linear filter. It can be the Zero Forcer (ZF) filter, which is described by
\begin{equation}
\mathbf{a}_{\textrm{ZF}}=\frac{\mathbf{H}[t]\mathbf{\hat{E}}_0[t]}{\mathbf{\hat{E}}_0^H[t]\mathbf{H}^H[t]\mathbf{H}[t]\mathbf{\hat{E}}_0[t]}.
\label{eq:crzf}
\end{equation}
Alternatively, the Minimum Mean Square Error (MMSE) filter is described by
\begin{equation}
\mathbf{a}_{\textrm{MMSE}}=\left(\mathbf{H}[t]\mathbf{\hat{E}}_0[t]\mathbf{\hat{E}}_0^H[t]\mathbf{H}^H[t]+\mathbf{R_w}\right)^{-1}\mathbf{H}[t]\mathbf{\hat{E}}_0[t].
\label{eq:crmmse}
\end{equation}

It is important to remark that the PSK estimation depends on the estimation of $\mathbf{\hat{E}}_0$ and thus, the errors of the estimators are propagated.

\begin{figure}[!ht]
\centering
\includegraphics[width=0.9\linewidth,clip=true]{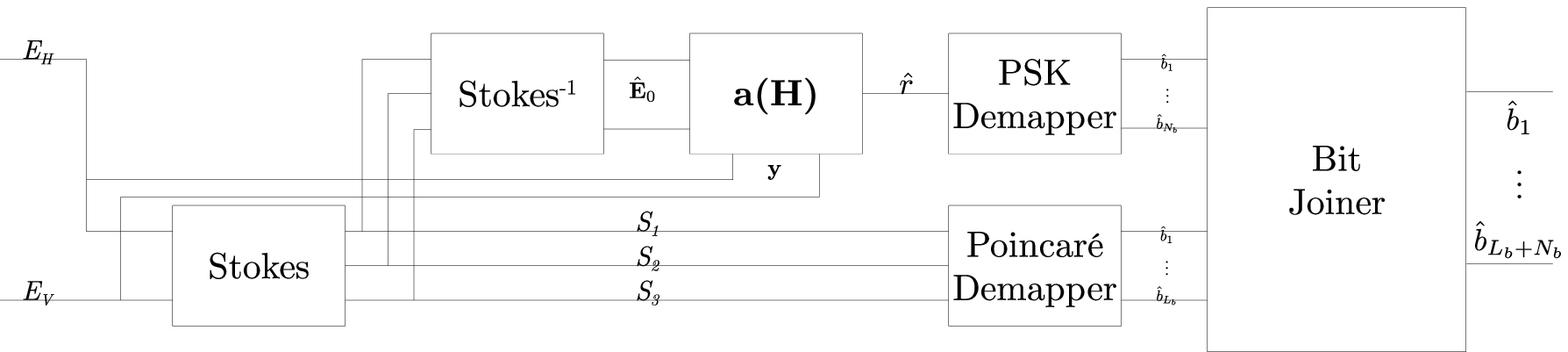}
\caption{Block diagram of 3D Polarized Modulation Cascade Receiver.}
\label{fig:3DPMod_diagramRXI}
\end{figure}

This receiver has lower computational complexity compared with the Joint Receiver. In particular, the computational complexity is the sum of each sub-receivers, i.e., $o(L)+o(N)$. Fig. \ref{fig:3DPMod_diagramRXI} illustrates the block diagram of this receiver. $\mathbf{a(H)}$ applies the filter operation described by \eqref{eq:crzf} or \eqref{eq:crmmse} and depends on the channel matrix $\mathbf{H}[t]$. In the simulation section this receiver is evaluated and compared with the optimal one; thus, concluding within which SNR range its performance is competitive.

\section{BER Analysis}
In \cite{Wei2012,Wei2013a}, the authors describe the instantaneous Symbol Error Rate (SER), as a function of several multidimensional integrals. However, these expressions do not provide closed-form solutions and are complex to compute. In \cite{Renzo2012}, the authors examine the same problem and propose a tight upper bound. 

We perform the BER analysis of the proposed 3D PMod scheme by means of pairwise error probability (PEP) and union bound, defined in \cite{Renzo2012} as
\begin{multline}
\textrm{BER}\leq\frac{1}{LN}\frac{1}{L_bN_b}\\\sum_{l=1}^L\sum_{l'=1}^L\sum_{n=1}^N\sum_{n'=1}^N\mathcal{D}\left(\left(l',n'\right)\rightarrow(l,n)\right)\textrm{PEP}\left(\left(l',n'\right)\rightarrow(l,n)\right),
\label{eq:BERall}
\end{multline}
where $\mathcal{D}\left(\left(l',n'\right)\rightarrow(l,n)\right)$ is the Hamming distance, i.e., the number of different bits between symbol defined by $\left(l',n'\right)$ and $(l,n)$. A very interesting aspect of \cite{Renzo2012} is the fact that the BER can be decoupled into three contributions. A symbol is decoded erroneously if 1) the polarization is estimated correctly but the initial phase is erroneous, 2) the initial phase estimation is correct but the estimated polarization state is erroneous, 3) neither the polarization state nor the initial phase are estimated correctly. Hence, the BER contributions are described as follows:
\begin{itemize}
	\item BER obtained by the distance between the symbols belonging to the same PSK constellation, i.e.,  $\left(\left(l,n'\right)\rightarrow(l,n)\right),\ \forall n'\neq n$.
	\item BER obtained by the distance between the symbols belonging to the PolSK constellation, i.e., $\left(\left(l',n\right)\rightarrow(l,n)\right),\ \forall l'\neq l$.
	\item BER obtained by the distance between the symbols belonging to the PolSK and PSK  $\left(\left(l',n'\right)\rightarrow(l,n)\right),\ \forall l'\neq l,\ \forall n'\neq n$.
\end{itemize}
Note that the aforementioned distances are referred to the I/Q plane, i.e., the signal after Stokes to Jones conversion.

A very interesting observation is that the first BER can be expressed in terms of exact error probability and is widely available in the literature, without incurring into bounding and obtain accurate results. This produces tighter union upper bound. Note that BER depends on the bit mapping. For the sake of homogeneity, we assume Gray bit mapping. Gray mapping is used vastly in the literature and it is proven that produces the lowest BER, as it maximizes the Hamming distance. Thus, \eqref{eq:BERall} can be described as the contribution of three terms:
\begin{equation}
\textrm{BER}\leq\textrm{BER}_{\textrm{Signal}}+\textrm{BER}_{\textrm{PolSK}}+\textrm{BER}_{\textrm{Joint}},
\end{equation}
where 
\begin{multline}
\textrm{BER}_{\textrm{Signal}}=\frac{N_b}{L_b+N_b}\textrm{BER}_{\textrm{PSK}}\\
\textrm{BER}_{\textrm{PolSK}}=\frac{1}{L}\frac{1}{L_b+N_b}\sum_{l=1}^L\sum_{l'=1}^L\mathcal{D}\left(l'\rightarrow l\right)Q\left(\sqrt{\frac{d_{l',l}^2}{2N_0}}\right)\\
\textrm{BER}_{\textrm{Joint}}=\frac{1}{LN}\frac{1}{L_b+N_b}\sum_{l=1}^L\sum_{n=1}^N\sum_{l'=1}^L\sum_{n'=1}^N\left(\mathcal{D}\left(l'\rightarrow l\right)\right.\\\left.+\mathcal{D}\left(n'\rightarrow n\right)\right)Q\left(\sqrt{\frac{d_{l',l,n',n}^2}{2N_0}}\right).
\end{multline}
The term $\textrm{BER}_{\textrm{PSK}}$ can be substituted by the exact BER expression, since it is known in the literature \cite{Proakis} and takes the following expression
\begin{multline}
\textrm{BER}_{\textrm{PSK}}=\frac{1}{N_b}\left(1\right.\\\left.-\frac{1}{2\pi}\int_{-\frac{\pi}{N}}^{\frac{\pi}{N}}e^{-\gamma\sin^2\theta}\int_{0}^{\infty}\nu e^{-\frac{\left(\nu-\sqrt{2\gamma}\cos\theta\right)^2}{2}}\dif\nu\dif\theta\right).
\end{multline}
The previous integral has a closed-form expression only in the case of $N=2$ and $N=4$. In these cases, the expression is reduced to
\begin{equation}
\textrm{BER}_{\textrm{PSK}}=Q\left(\sqrt{2\gamma}\right)
\end{equation}
for $N=2$ and
\begin{equation}
\textrm{BER}_{\textrm{PSK}}=Q\left(\sqrt{2\gamma}\right)\left(1-\frac{Q\left(\sqrt{2\gamma}\right)}{2}\right)
\end{equation}
for $N=4$, where $\gamma=\mathcal{E}/N_0$.

Before developing the expressions of the distances $d_{l',l}$ and $d_{l',l,n',n}$, for the sake of simplicity, we express the Stokes vector as a function of spherical coordinates $(\phi,\vartheta)$, where $\phi\in[0,2\pi]$ and $\vartheta\in[0,\pi]$. Hence,
\begin{equation}
\mathbf{S}_l\equiv\mathcal{E}\begin{pmatrix}1\\\cos\vartheta_l\\\sin\vartheta_l\cos\phi_l\\\sin\vartheta_l\sin\phi_l\end{pmatrix}
\end{equation}

Using \eqref{eq:Exfieldsph}, the generic distance $d_{l',l,n',n}$ in the Euclidean space is expressed as the norm of two arbitrary symbols $\|\mathbf{x}_{l',n'}-\mathbf{x}_{l,n}\|$. Thus,
\begin{multline}
d_{l',l,n',n}^2=2\mathcal{E}\left(1-\left(\cos\left(\Delta\xi-\frac{\Delta\phi}{2}\right)\cos\frac{\vartheta_{l'}}{2}\cos\frac{\vartheta_l}{2}\right.\right.\\\left.\left.+\cos\left(\Delta\xi+\frac{\Delta\phi}{2}\right)\sin\frac{\vartheta_{l'}}{2}\sin\frac{\vartheta_l}{2}\right)\right)
\label{eq:ds1s2}
\end{multline}
where $\Delta\xi=\xi_{n'}-\xi_n$, $\Delta\phi=\phi_{l'}-\phi_l$, $\xi_n$ is the PSK $n$ symbol, $(\phi_l,\vartheta_l)$ is the 3D PolSK $l$ symbol, composing the $\mathbf{x}_{l,n}$ 3D PMod symbol. This expression is the general version for an arbitrary pair of symbols.

The previous distance is further reduced if both symbols have the same PSK component, i.e., $\Delta\xi=0$. Then, \eqref{eq:ds1s2} is reduced to
\begin{equation}
d_{l',l}^2=2\mathcal{E}\left(1-\cos\left(\frac{\Delta\phi}{2}\right)\cos\left(\frac{\Delta\vartheta}{2}\right)\right),
\label{eq:mindistpolsk}
\end{equation}
where $\Delta\vartheta=\vartheta_{l'}-\vartheta_l$.

Note that if both symbols belong the same PolSK position, i.e., $\Delta\phi=0$ and $\vartheta_{l'}=\vartheta_l$, the distance expression \eqref{eq:ds1s2} is reduced to
\begin{equation}
d_{n',n}^2=2\mathcal{E}\left(1-\cos\Delta\xi\right)=4\mathcal{E}\sin^2\left(\frac{\pi\Delta n}{N}\right),
\label{eq:mindistpsk}
\end{equation}
where $\Delta n=n'-n$, which is equivalent to the well known PEP of PSK \cite{Proakis}.

However, the distance of PolSK cannot be studied analytically, since there is no closed-form expression on the symbols belonging the constellation (they are obtained numerically \cite{Sloane}). Despite this, we are able to compute this distance numerically for different modulation orders. 

The packing problem is not a new problem. Essentially, it aims at finding an answer to the question \emph{How $n$ points should be placed on a sphere surface in such a way that the minimum distance between them is maximized?} This problem is known as \emph{Tammes Problem} \cite{Tammes1930,Kottwitz1991}. Unfortunately, there is no closed solution, although there are known solutions for a small number of points. For instance:
\begin{itemize}
	\item $L=1$: the solution is trivial.
	\item $L=2$: points at the poles.
	\item $L=3$: points at the equator separated $120$ degrees apart.
	\item $L=4$: vertices of a regular tetrahedron.
	\item ...
\end{itemize}

In particular, for $L=2,4,8,16$ simple solutions can be found. The z-axis is sliced in few levels with symmetry in the equator and the points are located equispaced in each slice.

The BER analysis can be also performed in terms of minimum distance. In the presence of AWGN, the overall performance of the system is mainly described by the minimum distance. Thus, we compare the minimum distance using different $L\times N$ combinations for several spectral efficiencies. Tables \ref{tab:mindist2}, \ref{tab:mindist3}, \ref{tab:mindist4}, \ref{tab:mindist5}, \ref{tab:mindist6}, \ref{tab:mindist7} and \ref{tab:mindist8} summarize the minimum distance of different schemes for a fixed spectral efficiencies of $2, 3, 4, 5, 6, 7, 8$ bits. The results in boldface denote the mode with the maximum minimum distance. For the sake of clarity, the results are expressed in numeric form instead of using trigonometric functions and fractions.

\begin{figure}[!ht]
	\parbox{.45\linewidth}{
		\centering
		\includegraphics[width=0.95\linewidth,clip=true]{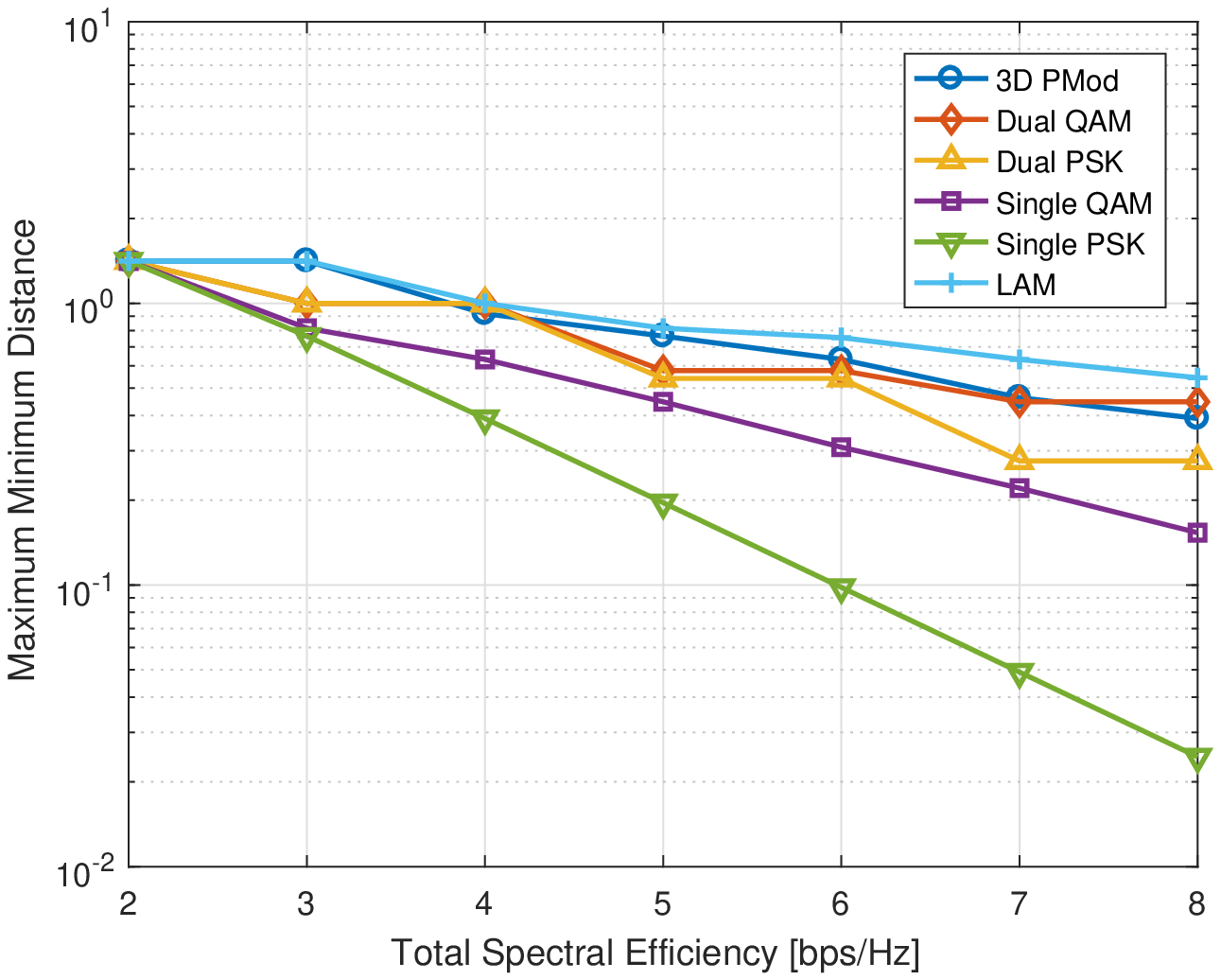}
		\caption{Maximum minimum distance for different spectral efficiencies.}
		\label{fig:minDist}
	}
	\hfill
	\parbox{.45\linewidth}{
		\centering
		\includegraphics[width=0.95\linewidth,clip=true]{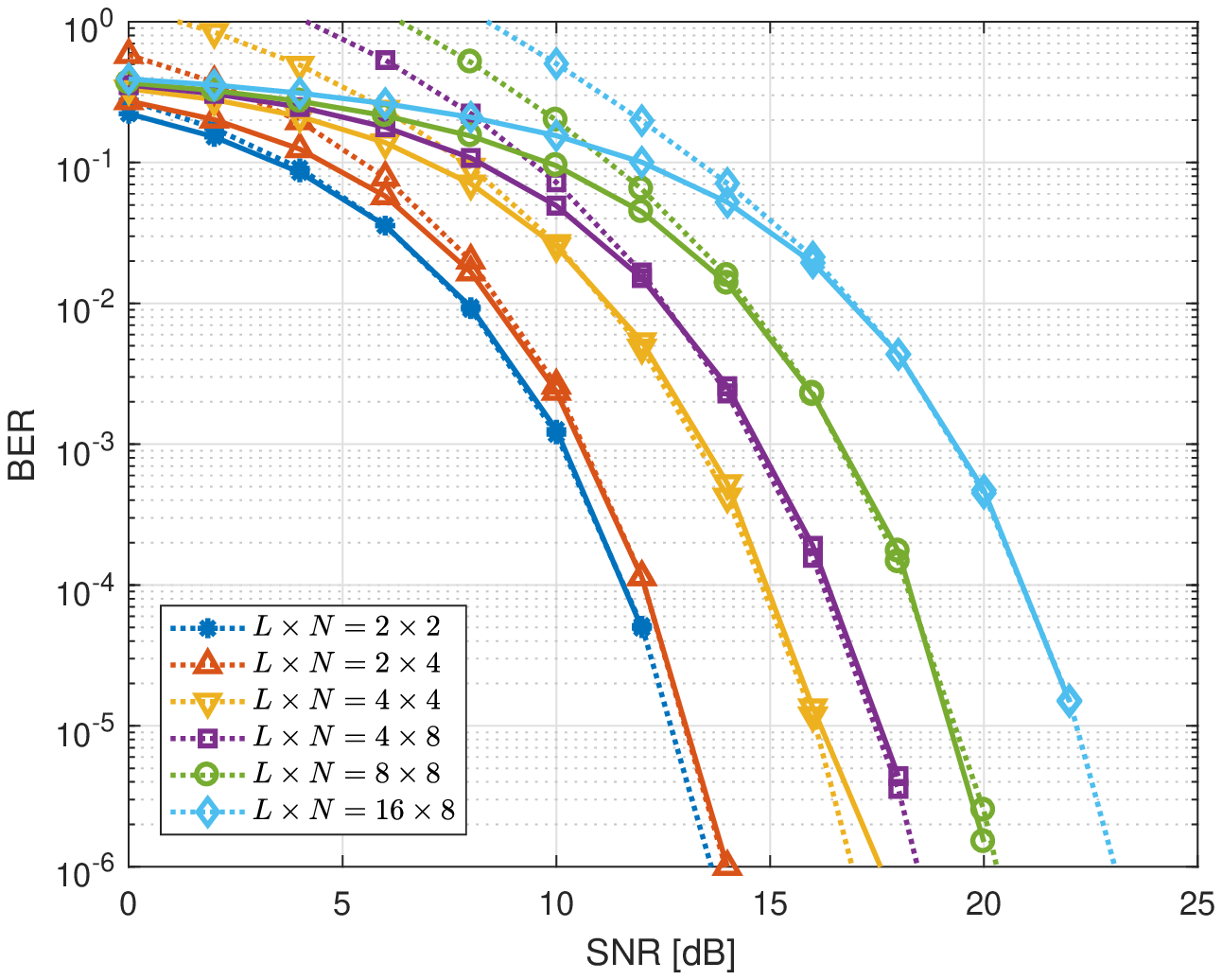}
		\caption{BER of 3D PMod for the different considered modes. Solid lines are obtained via Monte Carlo simulations. Dashed lines correspond to the Union Bound \eqref{eq:BERall}.}
		\label{fig:res_BER_all}
	}
\end{figure}

Fig. \ref{fig:minDist} depicts the maximum minimum distance for different spectral efficiencies compared with common schemes, such as Dual QAM, Dual PSK, Single QAM, Single PSK and LAM. These schemes are described as follows:

\begin{itemize}
	\item 3D PMod: the proposed scheme described in \eqref{eq:Exfield} and \eqref{eq:Exfieldsph}.
	\item Dual QAM scheme conveys $L$-QAM and $N$-QAM constellations in each horizontal or vertical polarization.
	\item Dual PSK scheme conveys $L$-PSK and $N$-PSK constellations in each horizontal or vertical polarization.
	\item Single QAM conveys an $L\times N$-QAM constellation using a single polarization.
	\item Single PSK conveys an $L\times N$-PSK constellation using a single polarization.
	\item LAM conveys an $L\times N$-LAM constellation using both polarizations.
\end{itemize}
Note that Fig. \ref{fig:minDist} is not obtained via Monte Carlo simulations since it is produced analytically by computing the minimum distance between all symbols in the constellation.

Examining the respective tables, we can conclude the following aspects:
\begin{itemize}
	\item The minimum distance of $L\times N$ of 3D PMod is determined by the 3D PolSK contribution if $L>N$ or by the PSK contribution if $L<N$. In the case  where $L=N$ both contributions are mixed.
	\item In the cases of Dual QAM and Dual PSK, the minimum distance is determined by the highest modulation order. Thus, it is equivalent to use the minimum distance of the constellations $\max(L,N)$-QAM/PSK, respectively. This is the reason of the flat areas in Fig. \ref{fig:minDist}.
	\item The minimum distance of Single QAM and Single PSK can be computed using the known formulas for an $L\times N$-QAM/PSK constellations \cite{Proakis}.
	\item For low modulation orders (below than $8$ bits) 3D PMod achieves the maximum performance, below LAM, except $4\times 4$. The reason for this is that the Dual QPSK double polarization has higher minimum distance.
	\item For asymmetric $L\times N$ schemes, 3D PMod achieves higher performance if the deviation of $L$ and $N$ is lower. For instance, $4\times8$ and $8\times4$ have higher minimum distance compared with $2\times16$ and $16\times2$ for the same spectral efficiency. This is also valid for Dual QAM/PSK. This is because the minimum distance is constraint by the highest order $L$ or $N$.
\end{itemize}

It is worth to mention that LAM is designed in such a way that the minimum distance is maximized using both polarizations. The constellation can be envisaged as a 4D constellation, where the points are placed in a hypercube. This constellation is often referred as the optimal, since it achieves the highest mutual information \cite{Mazzali2016}. However, LAM constellations present major drawbacks compared to 3D PMod:
\begin{itemize}
	\item The benefits of LAM are observable for $L\times N>=64$. The benefits of 3D PMod are observable for $L\times N < 256$. Thus, 3D PMod is specially indicated for low and medium SNR regimes, whereas LAM is more focused for high SNR regimes.
	\item The computational complexity is unaffordable since it requires to implement the ML receiver as well during the constellation design.
	\item The bit mapping is not trivial. Gray mapping cannot be always applied and, hence, the BER performance is not always the optimal.
	\item LAM design is based on spherical cuts of a lattice structure. Hence, the peak-to-average ratio impact is not negligible. 3D PMod has a constant joint envelope, i.e., $\Expect\{\|\mathbf{x}\|^2\}=\|\mathbf{x}\|^2=\mathcal{E}$. 
	\item LAM does not allow symbol multiplexing nor codeword, whereas 3D PMod does.
	\item LAM does not support differentiated modulation order schemes. Both polarizations constitute a single supersymbol.
\end{itemize}

In \cite{Henarejos2015} and \cite{Henarejos2015a} we describe the communication system of 2D PMod. This is a particular case of 3D PMod, where $L=2$. In detail, we constraint it to V/H or RHCP/LHCP. It is clear that when H/V is used, only one channel is activated, corresponding to the horizontal or vertical polarization. Thus, the BER analysis described in \cite{Renzo2012} can be applied straightforwardly. 

In terms of minimum distance, it is determined by the PSK constellation and takes the expression of \eqref{eq:mindistpsk}. Examining tables \ref{tab:mindist2} it is interesting to see that 2D PMod achieves the maximum minimum distance for $2\times 2$ and $2\times 4$, i.e., orthogonal polarization + BPSK/QPSK constellations. For higher spectral efficiencies, modes with $L>2$ achieve higher minimum distance. This is particularly interesting since we demonstrate that 2D PMod obtains an appreciable performance for low modulation order schemes. 

\section{Results}
In this section we discuss the results obtained when 3D PMod is used. We implement the system described by Fig. \ref{fig:3DPMod_diagram} using different values of $L$ and $N$. In this figure, $\textrm{Bit Splitter}$ splits the incoming bits into two groups of $N_b$ and $L_b$ bits; $\textrm{PSK Mapper}$ and $e^j$ perform the PSK mapping; $\textrm{Poincar\'e Mapper}$ maps the incoming bits into the Poincar\'e sphere by using Tables \ref{tab:3dsym2}, \ref{tab:3dsym4}, \ref{tab:3dsym8} and \ref{tab:3dsym16} for $L=2,4,8,16$; $\textrm{Stokes}^{-1}$ performs the conversion from Stokes to Jones vector described by \eqref{eq:stokes2Esph}.

We also implemented the Joint Receiver and Cascade Receiver, described in the previous sections. All symbols are encoded using Gray coding and all results are obtained with AWGN channel. The transmission power is normalized to $0$ dBW and the timing and phase synchronization is assumed perfect. The Cascade Receiver uses the MMSE filter expression \eqref{eq:crmmse}. Based on the minimum distance analysis, we evaluate the following $L\times N$ modes to cover spectral efficiencies from $2$ to $7$ bps/Hz: $2\times 2$, $2\times 4$, $4\times 4$, $4\times 8$, $8\times 8$, $16\times 8$. 

We first perform an analysis by comparing both receivers, and depict the individual and joint BER. Although we do not depict the throughput, it is obtained by counting the number of symbols decoded successfully multiplied by the number of bits carried by the symbol. This is equivalent to
\begin{equation}
\textrm{Throughput}=\left(L_b+N_b\right)\left(1-\textrm{SER}\right)
\end{equation}
where $\textrm{SER}$ is the symbol error rate (SER). Note that SER can be computed using the XOR operator as follows
\begin{equation}
\begin{split}
\textrm{BER}&=\frac{1}{L_b+N_b}\sum_{n=1}^{L_b+N_b}\left|b_n-\hat{b}_n\right|\\
\textrm{SER}&=\biglor_{n=1}^{L_b+N_b}b_n\oplus\hat{b}_n 
\end{split}
\end{equation}
where $\biglor_{n=1}^Nx_n=x_1\lor\dots\lor x_N$ performs the logic OR operation and $\oplus$ is the logic XOR operation.

\subsection{Comparison of Classes of Receivers}

\begin{figure}[!ht]
	\centering
	\subfloat[][SE $2$ bps/Hz\\$L\times N=2\times 2$]{\includegraphics[width=0.45\linewidth,clip=true]{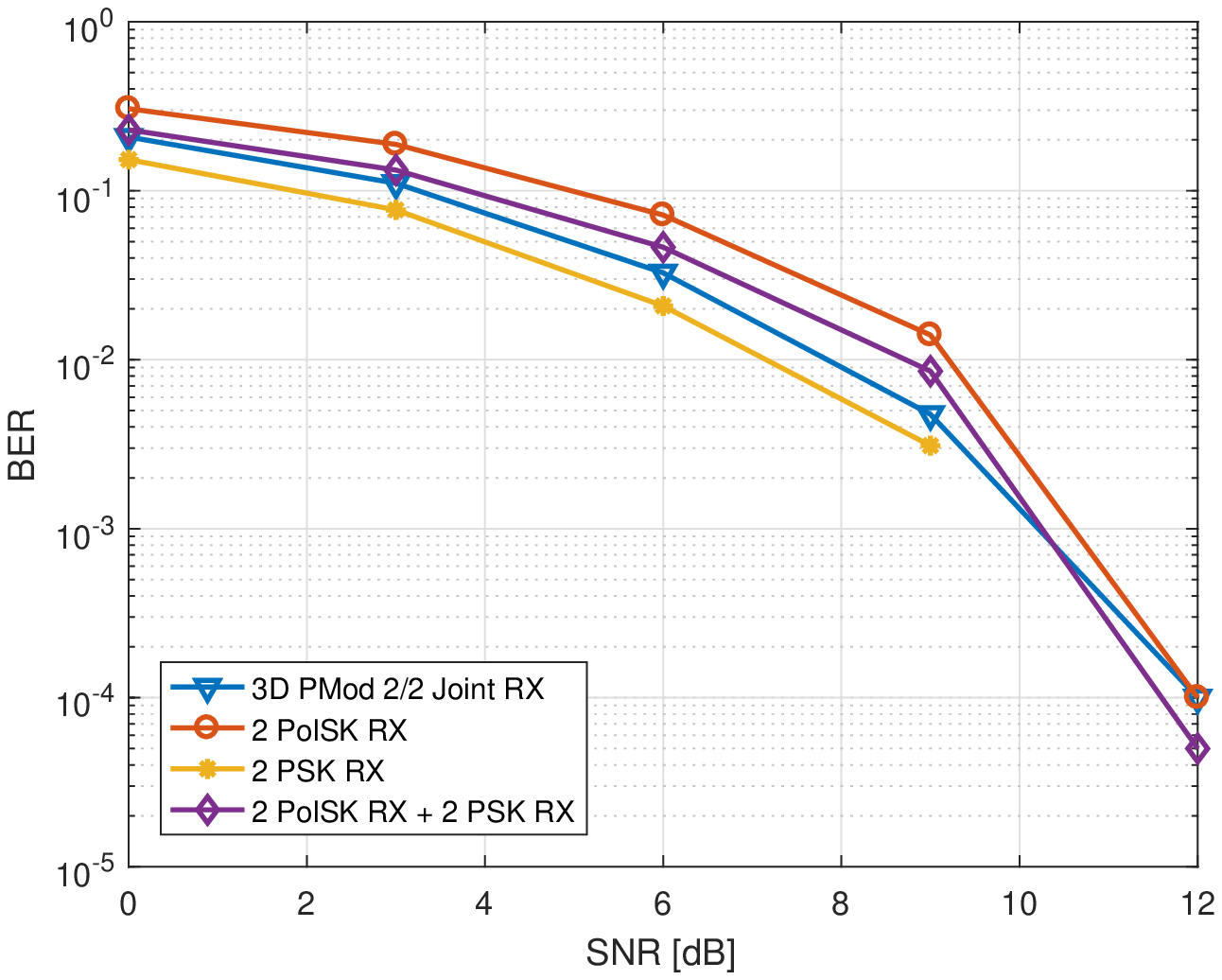}\label{fig:ber2x2_rx}}\hfill
	\subfloat[][SE $3$ bps/Hz\\$L\times N=2\times 4$]{\includegraphics[width=0.45\linewidth,clip=true]{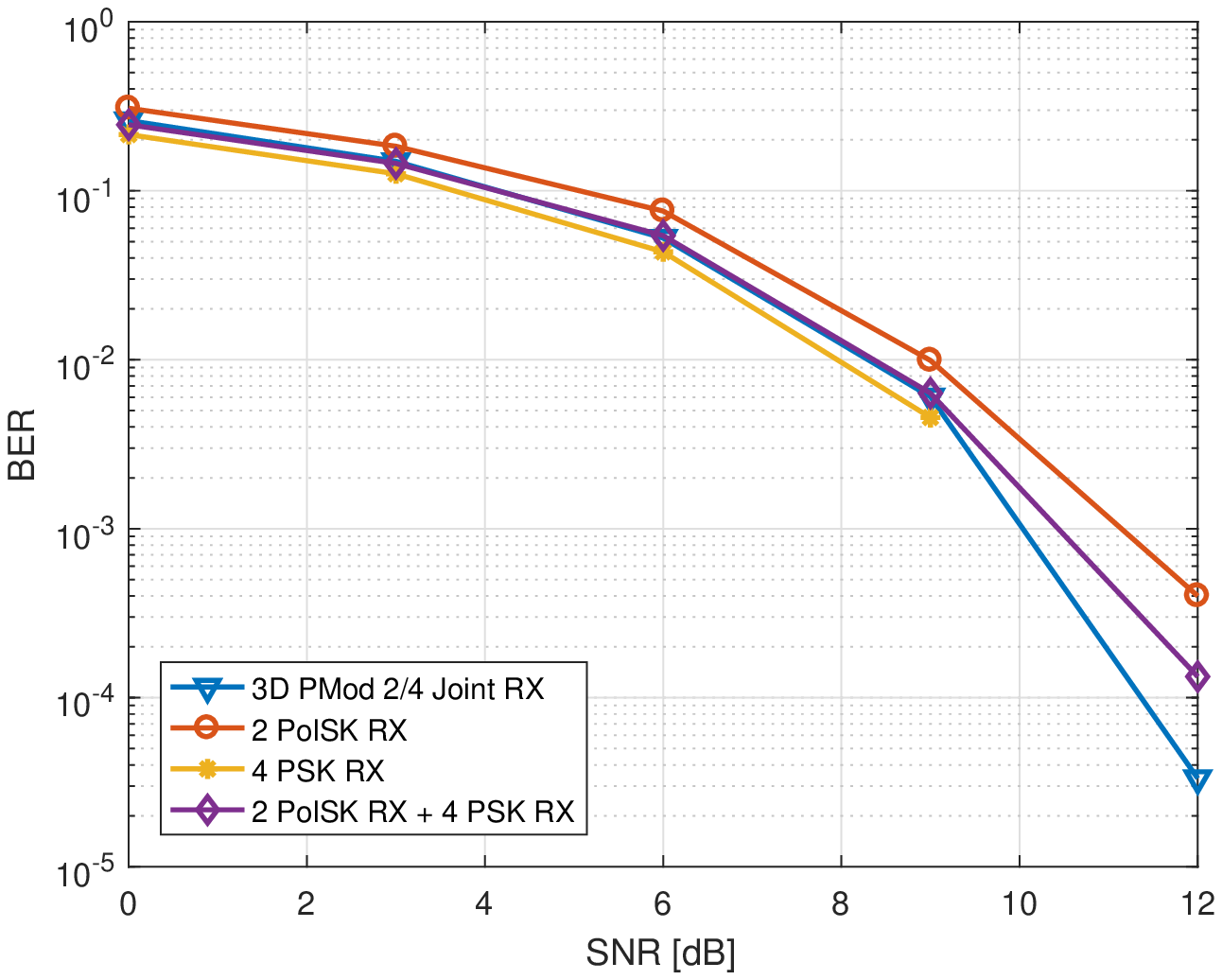}\label{fig:ber2x4_rx}}\hfill
	\subfloat[][SE $4$ bps/Hz\\$L\times N=4\times 4$]{\includegraphics[width=0.45\linewidth,clip=true]{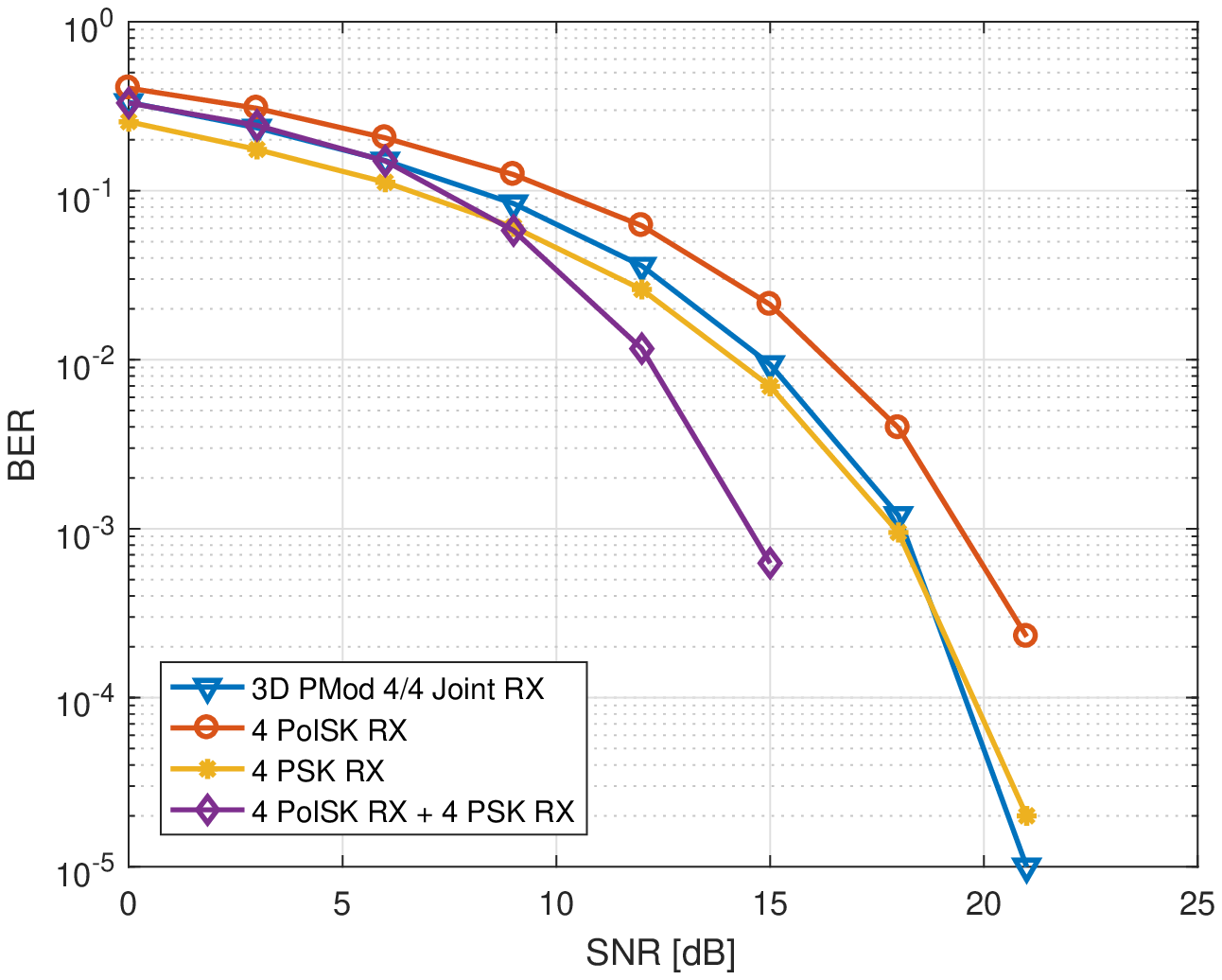}\label{fig:ber4x4_rx}}\hfill
	\subfloat[][SE $5$ bps/Hz\\$L\times N=4\times 8$]{\includegraphics[width=0.45\linewidth,clip=true]{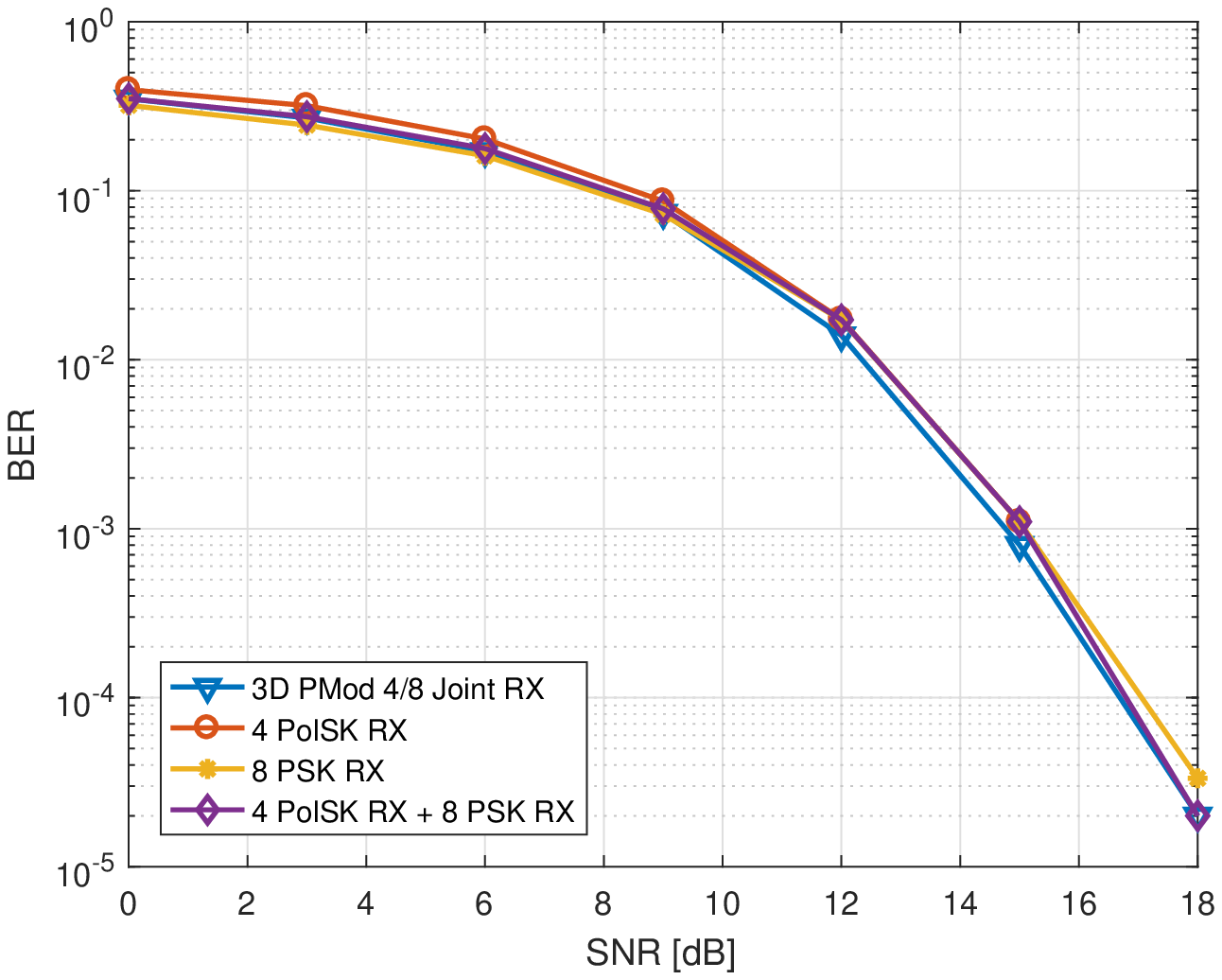}\label{fig:ber4x8_rx}}\hfill
	\subfloat[][SE $6$ bps/Hz\\$L\times N=8\times 8$]{\includegraphics[width=0.45\linewidth,clip=true]{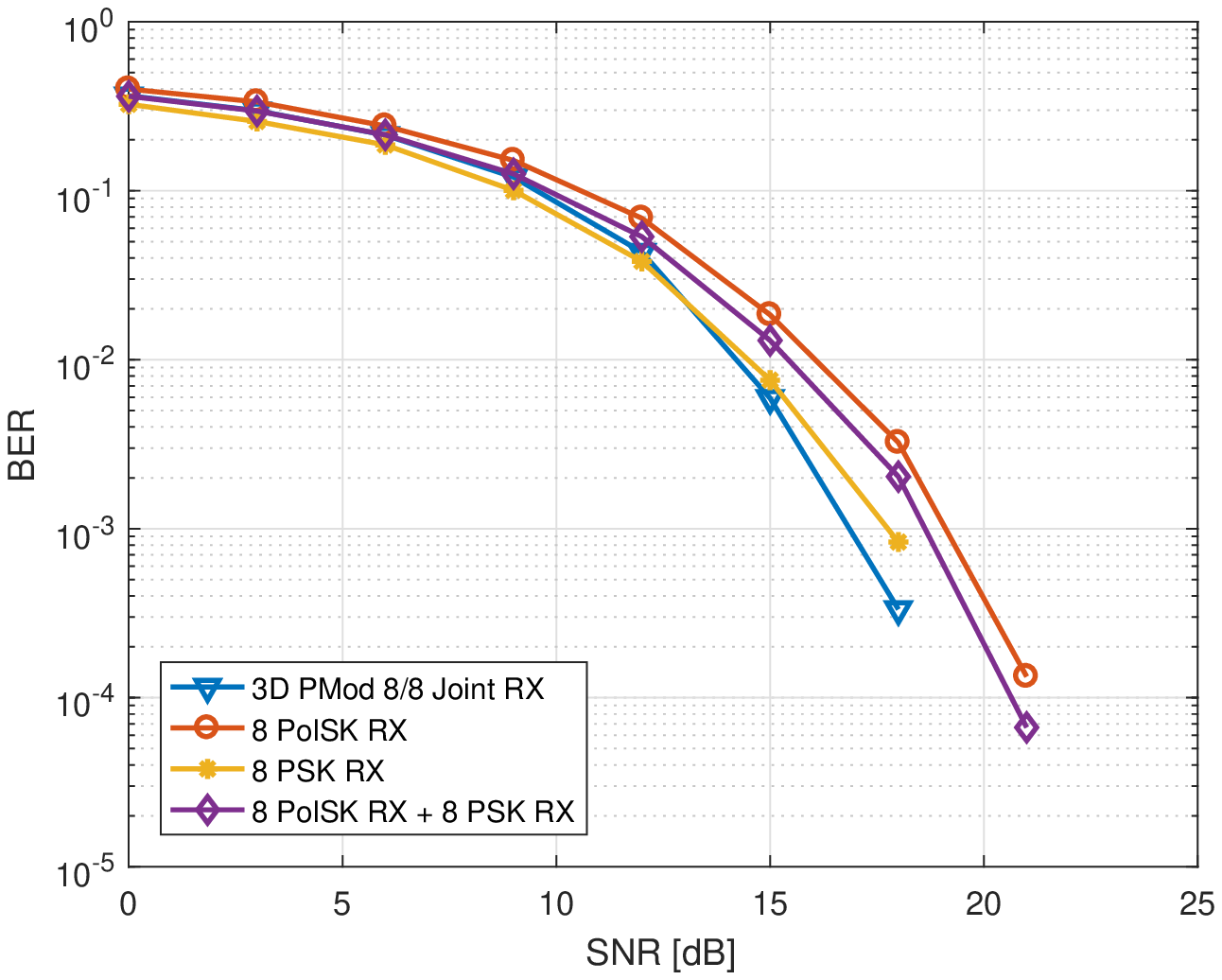}\label{fig:ber8x8_rx}}\hfill
	\subfloat[][SE $7$ bps/Hz\\$L\times N=16\times 8$]{\includegraphics[width=0.45\linewidth,clip=true]{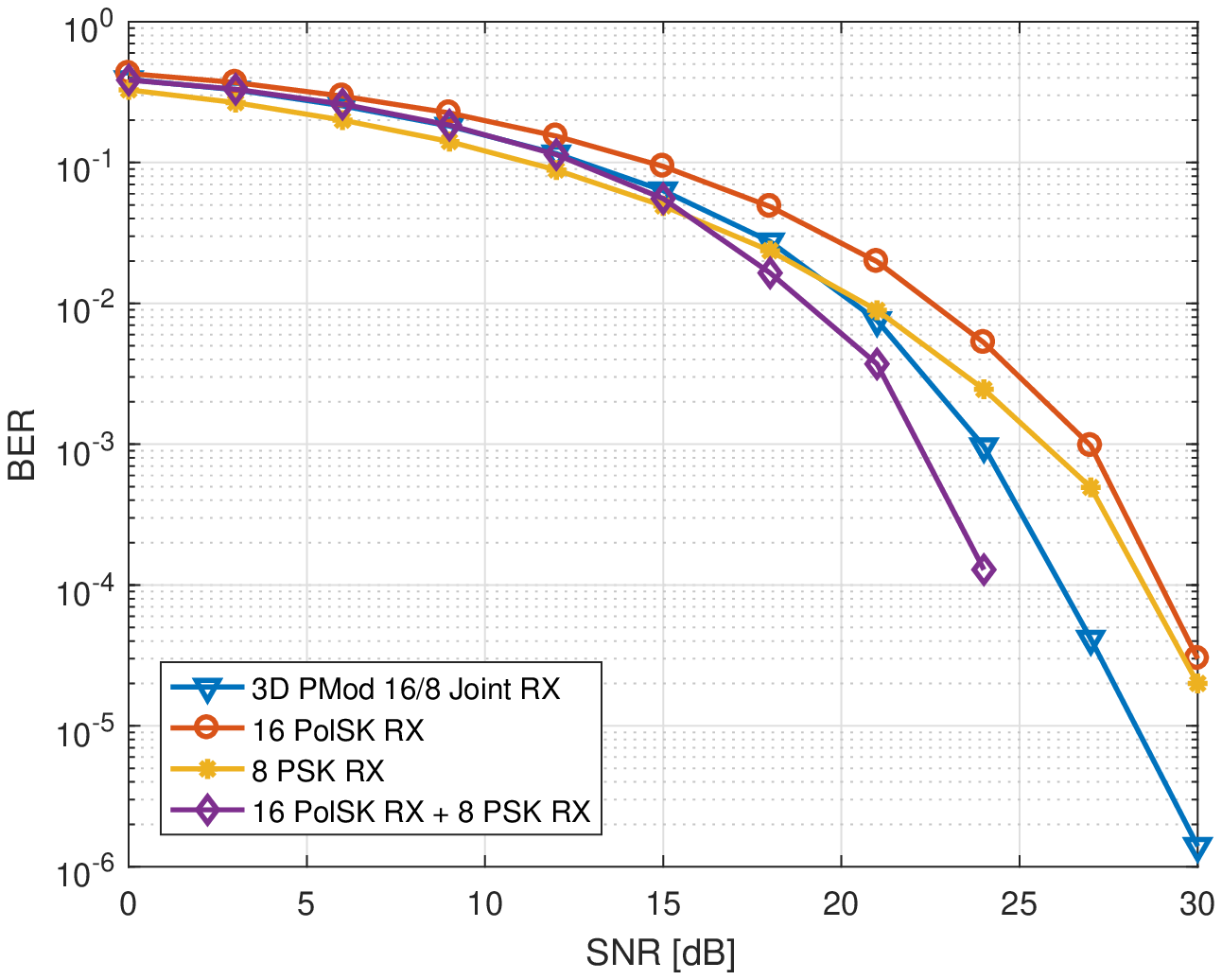}\label{fig:ber16x8_rx}}\hfill   
	\caption{Comparison of the BER of 3D Polarized Modulation for different classes of receivers. The combined BER from the Cascade sub-receivers is weighted by the number of bits carried by each modulation.}
	\label{fig:ber_3DPMod_rx}
\end{figure}

In this section we compare the performance achieved by each class of receivers. Fig. \ref{fig:ber_3DPMod_rx} illustrates the BER obtained by the different receivers. BERs labelled as \emph{Joint RX}, \emph{PolSK RX} and \emph{PSK RX} are obtained by using the Joint receiver and the Cascade PolSK and PSK sub-receivers. The BER labelled as \emph{PolSK RX + PSK RX} is obtained weighting the BERs of \emph{PolSK RX} and \emph{PSK RX} by the number of bits carried by each one. This figure shows that the Joint receiver outperforms the other schemes. Whilst the Joint receiver obtains lower BER at expenses of a higher computational complexity, the Cascade Receiver reduces drastically the computational complexity of the receiver at the expense of increasing the BER.


\subsection{Comparison of Different Modulation Orders}
To compare the different modulation orders in terms of BER, we use the Joint receiver as an optimal benchmark reference to compute the performance. Fig. \ref{fig:res_BER_all} depicts the BER of 3D PMod using different constellations. The Union Bound for each curve is also depicted in order to observe tight results. As expected, as the spectral efficiency increases, the minimum distance decreases, and therefore, the BER increases.


\subsection{Comparison with Other 3D Constellations}
\begin{figure}[!ht]
	\centering
	\subfloat[][$L=4$]{\includegraphics[width=0.45\linewidth,clip=true]{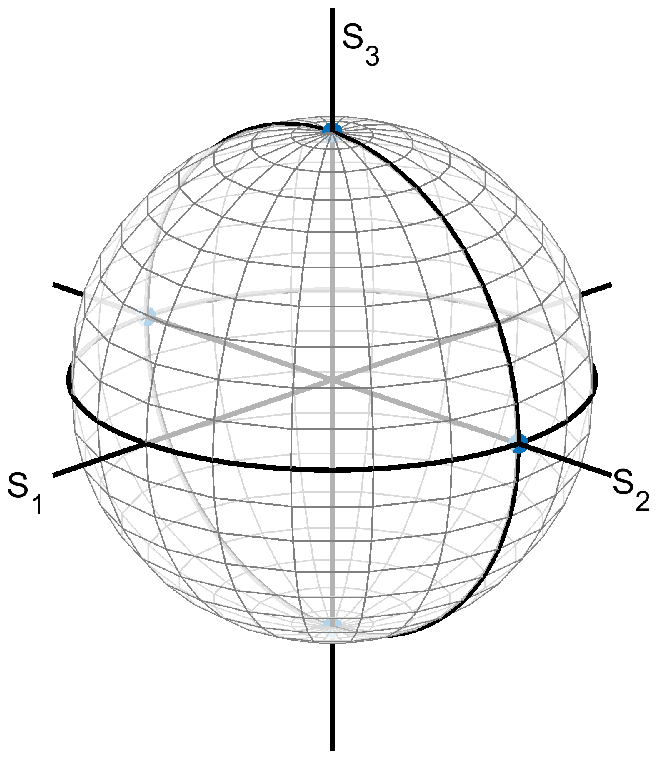}\label{fig:sphere4_Wei}}\hfill
	\subfloat[][$L=8$]{\includegraphics[width=0.45\linewidth,clip=true]{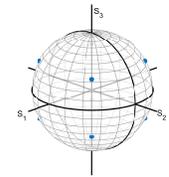}\label{fig:sphere8_Wei}}\hfill
	\caption{Constellations designed similarly to \cite{Wei2012}.}
	\label{fig:sphere_Wei}
\end{figure}

In this section we compare our approach by using the maximum minimum distance described by the Tammes problem with the constellations proposed in \cite{Wei2012}. Although the constellations in \cite{Wei2012} are not detailed, we designed constellations similar to them. Fig. \ref{fig:sphere_Wei} depicts the constellations similar to that in \cite{Wei2012}.

\begin{figure}[!ht]
	\centering
	\subfloat[][$L\times N=4\times 4$]{\includegraphics[width=0.45\linewidth,clip=true]{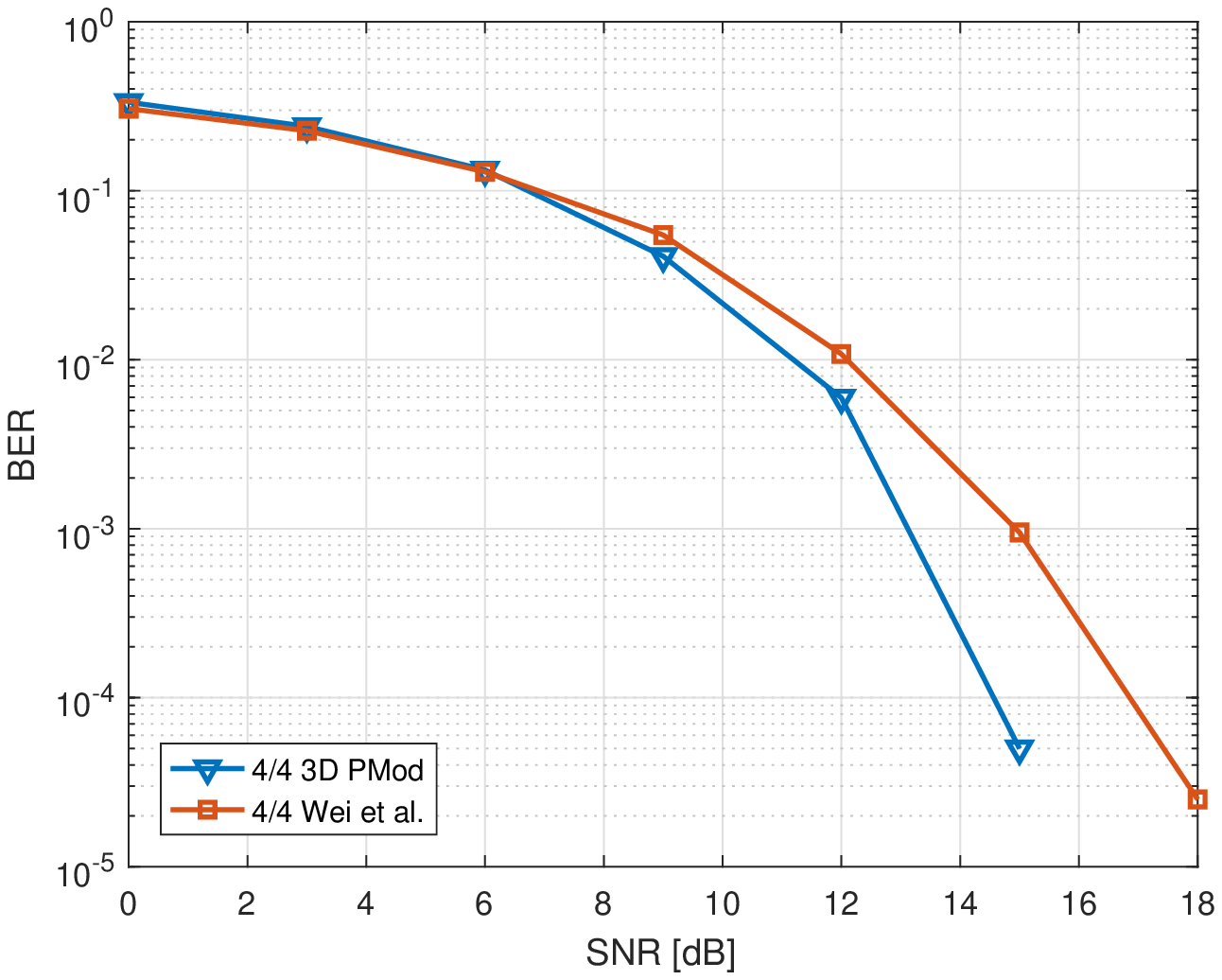}\label{fig:res_BER_4_4_Wei}}\hfill
	\subfloat[][$L\times N=8\times 4$]{\includegraphics[width=0.45\linewidth,clip=true]{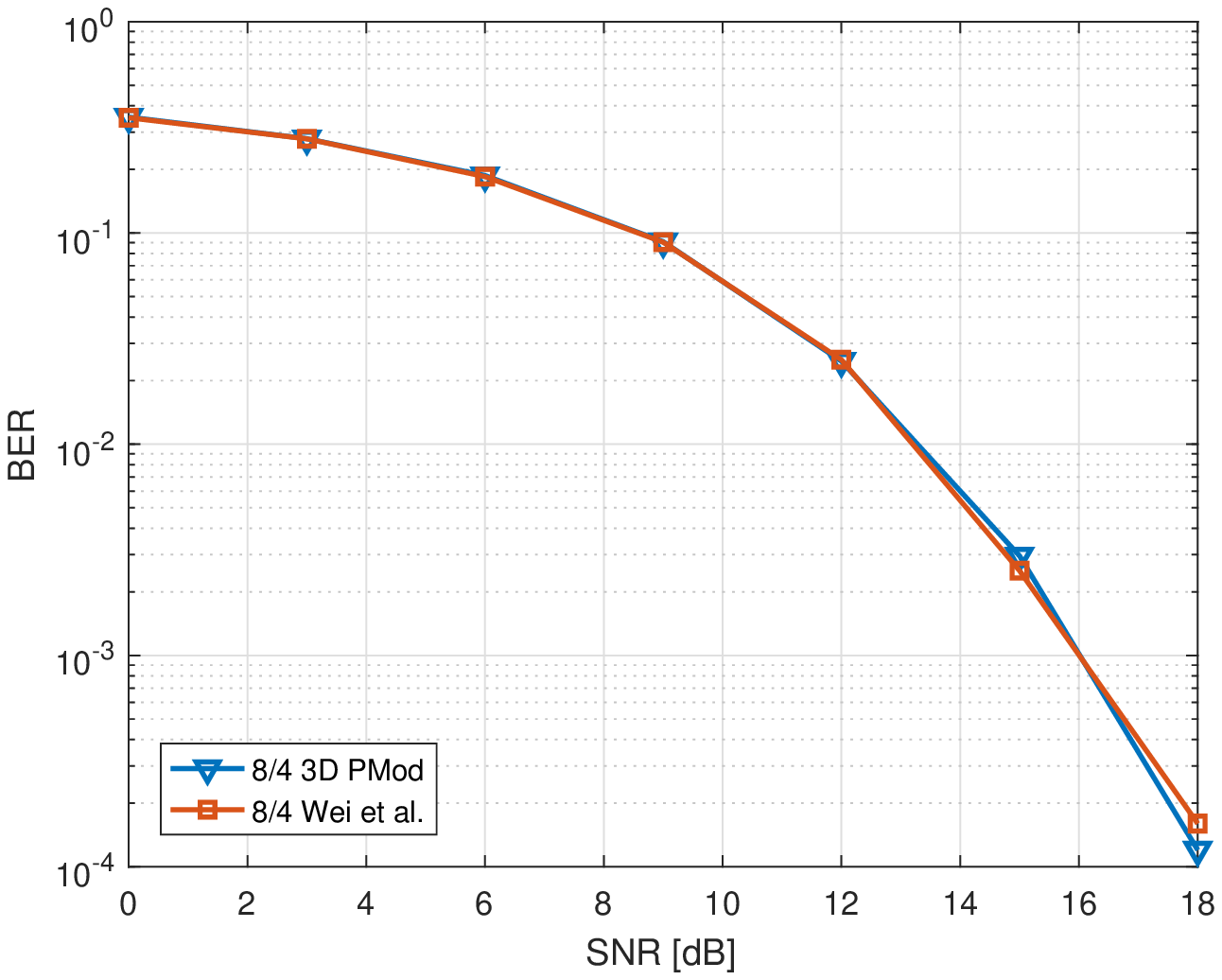}\label{fig:res_BER_8_4_Wei}}\hfill
	\caption{Comparison between our approach and \cite{Wei2012}.}
	\label{fig:BER_Wei}
\end{figure}

Fig. \ref{fig:BER_Wei} compares the BER of our approach and \cite{Wei2012}. As we remarked in the introduction, our approach uses the maximum minimum distance to obtain the lowest BER, whereas \cite{Wei2012} places 2D constellations equally spaced and regularly distributed along the x-axis. Therefore, as expected, our approach outperforms \cite{Wei2012} in terms of BER due to the criteria employed to design the constellation. Note that the maximum minimum distance can be compared by inspecting \ref{fig:expoincare} and \ref{fig:sphere_Wei}.

\subsection{Comparison with Other Existing Schemes}

In this section we compare the performance of the proposed schemes with other existing schemes. In the following figures, all schemes have the same spectral efficiency. We recall that, in the case of asymmetric sizes ($L\neq N$), Dual QAM and Dual PSK convey an $L$-symbol constellation through horizontal polarization and $N$-symbol constellation through vertical polarization. In the case of Single QAM and Single PSK, an $L\times N$-symbol constellation is conveyed through the horizontal polarization. Finally, in the case of LAM, an $L\times N$-symbol constellation is conveyed using both polarizations.

\begin{figure}[!ht]
	\centering
	\subfloat[][SE $2$ bps/Hz\\$L\times N=2\times 2$]{\includegraphics[width=0.45\linewidth,clip=true]{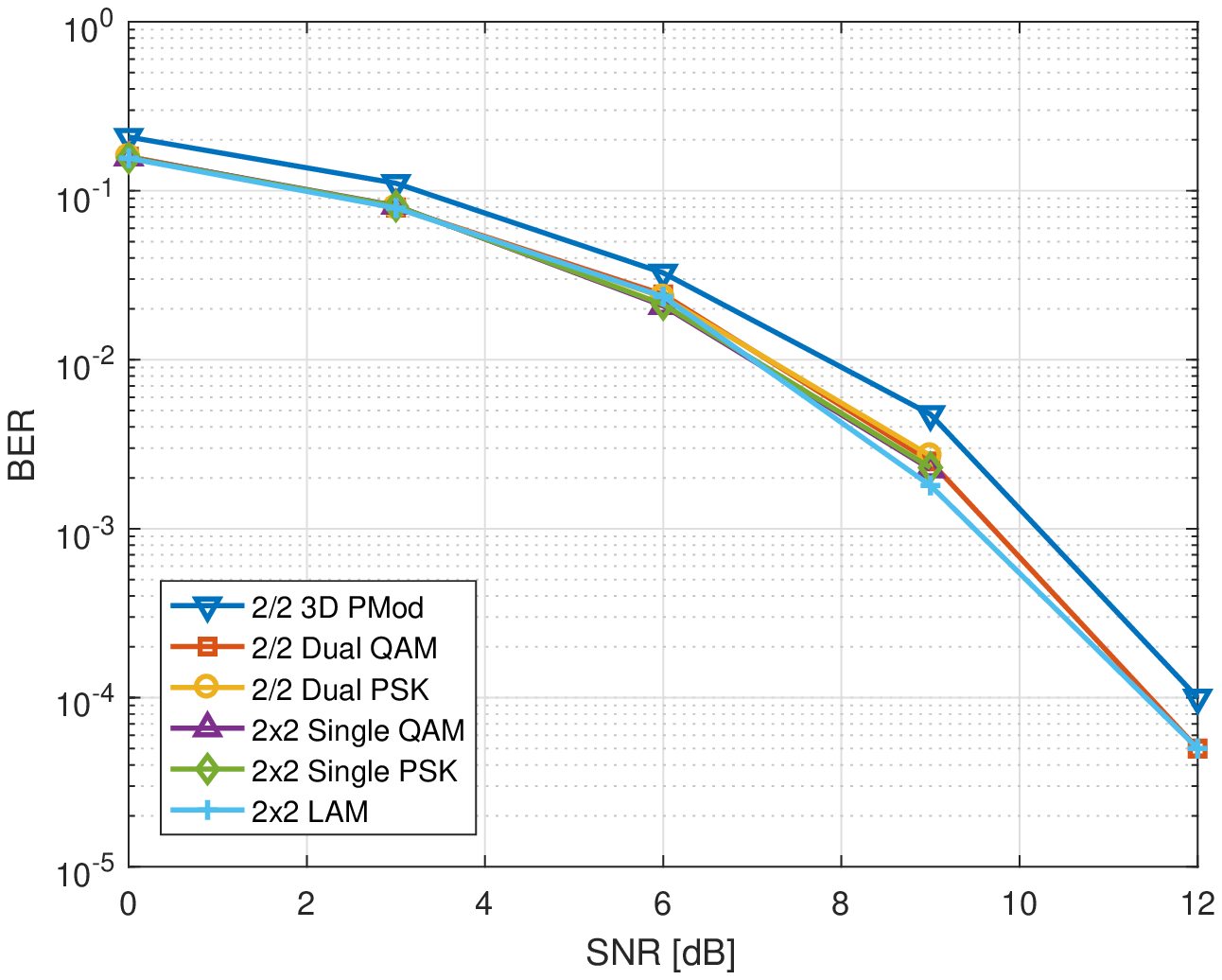}\label{fig:ber2x2}}\hfill
	\subfloat[][SE $3$ bps/Hz\\$L\times N=2\times 4$]{\includegraphics[width=0.45\linewidth,clip=true]{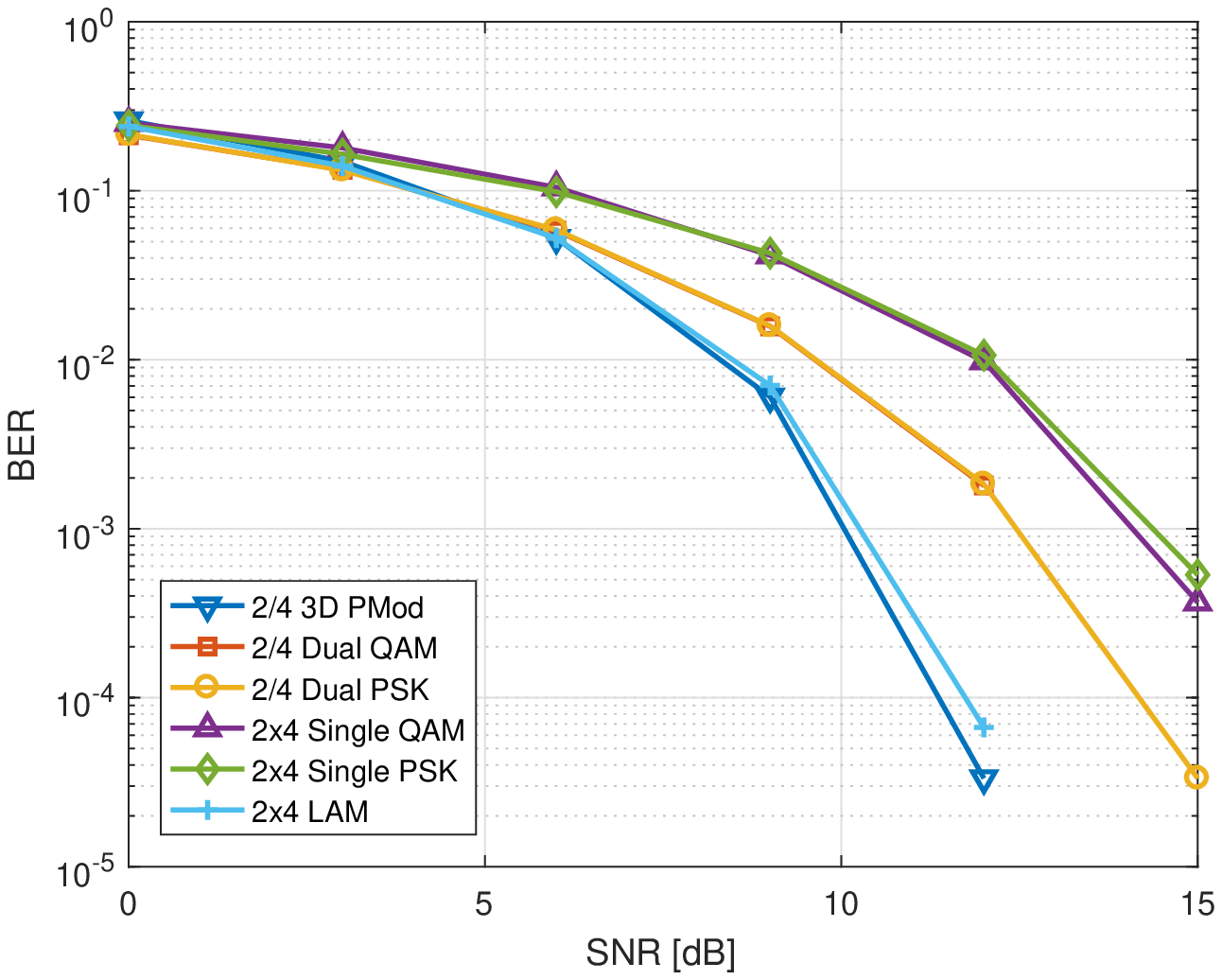}\label{fig:ber2x4}}\hfill
	\subfloat[][SE $4$ bps/Hz\\$L\times N=4\times 4$]{\includegraphics[width=0.45\linewidth,clip=true]{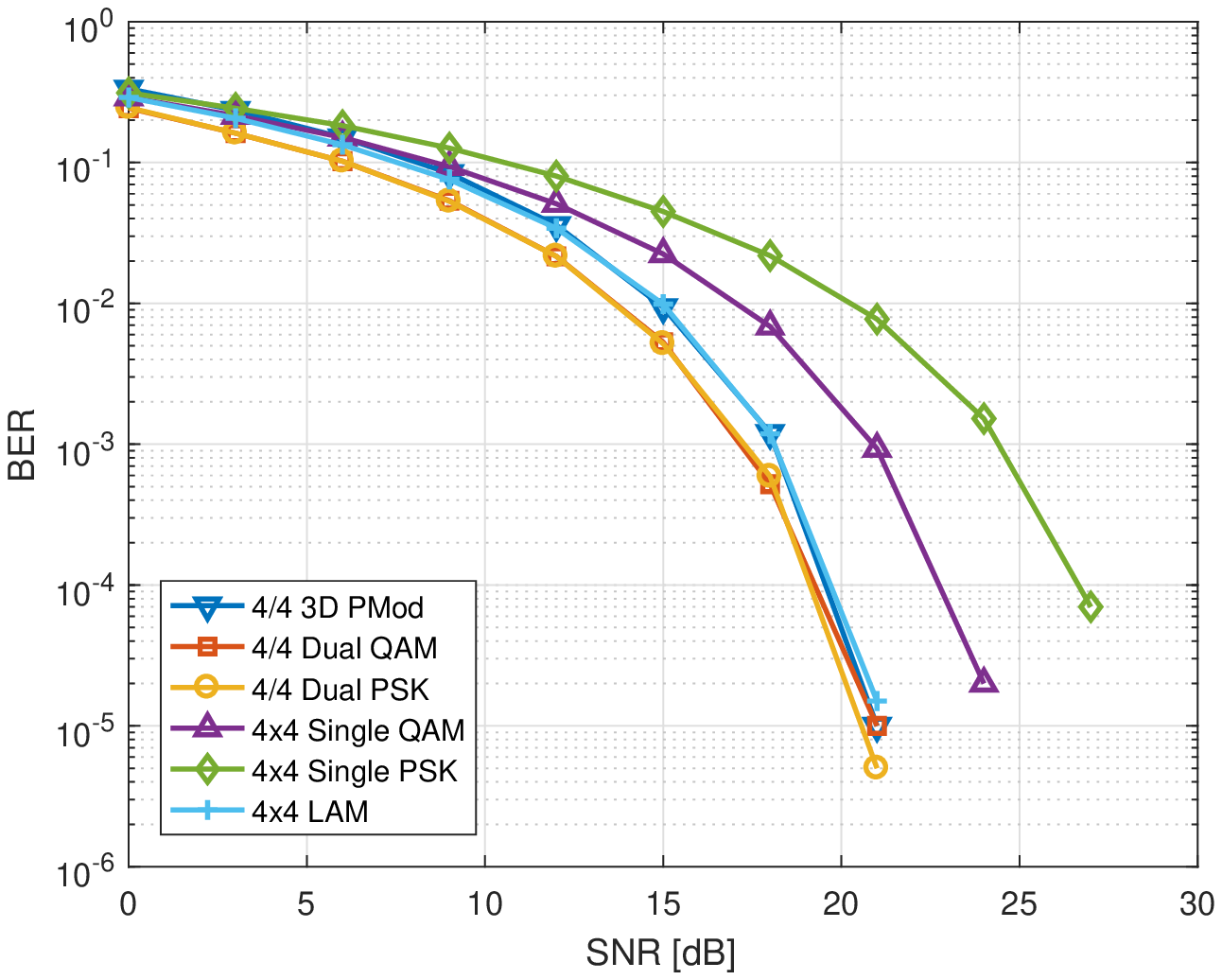}\label{fig:ber4x4}}\hfill
	\subfloat[][SE $5$ bps/Hz\\$L\times N=4\times 8$]{\includegraphics[width=0.45\linewidth,clip=true]{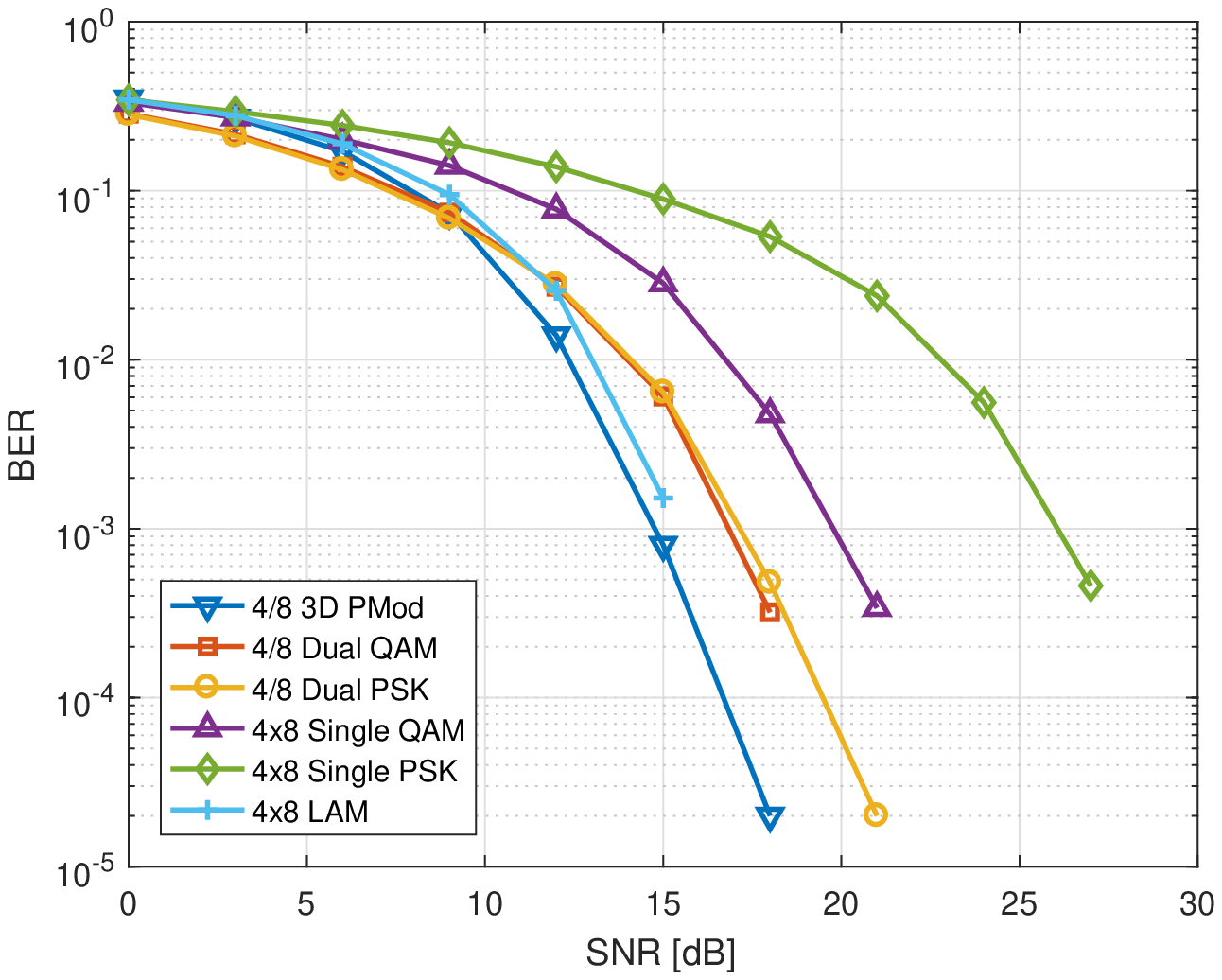}\label{fig:ber4x8}}\hfill
	\subfloat[][SE $6$ bps/Hz\\$L\times N=8\times 8$]{\includegraphics[width=0.45\linewidth,clip=true]{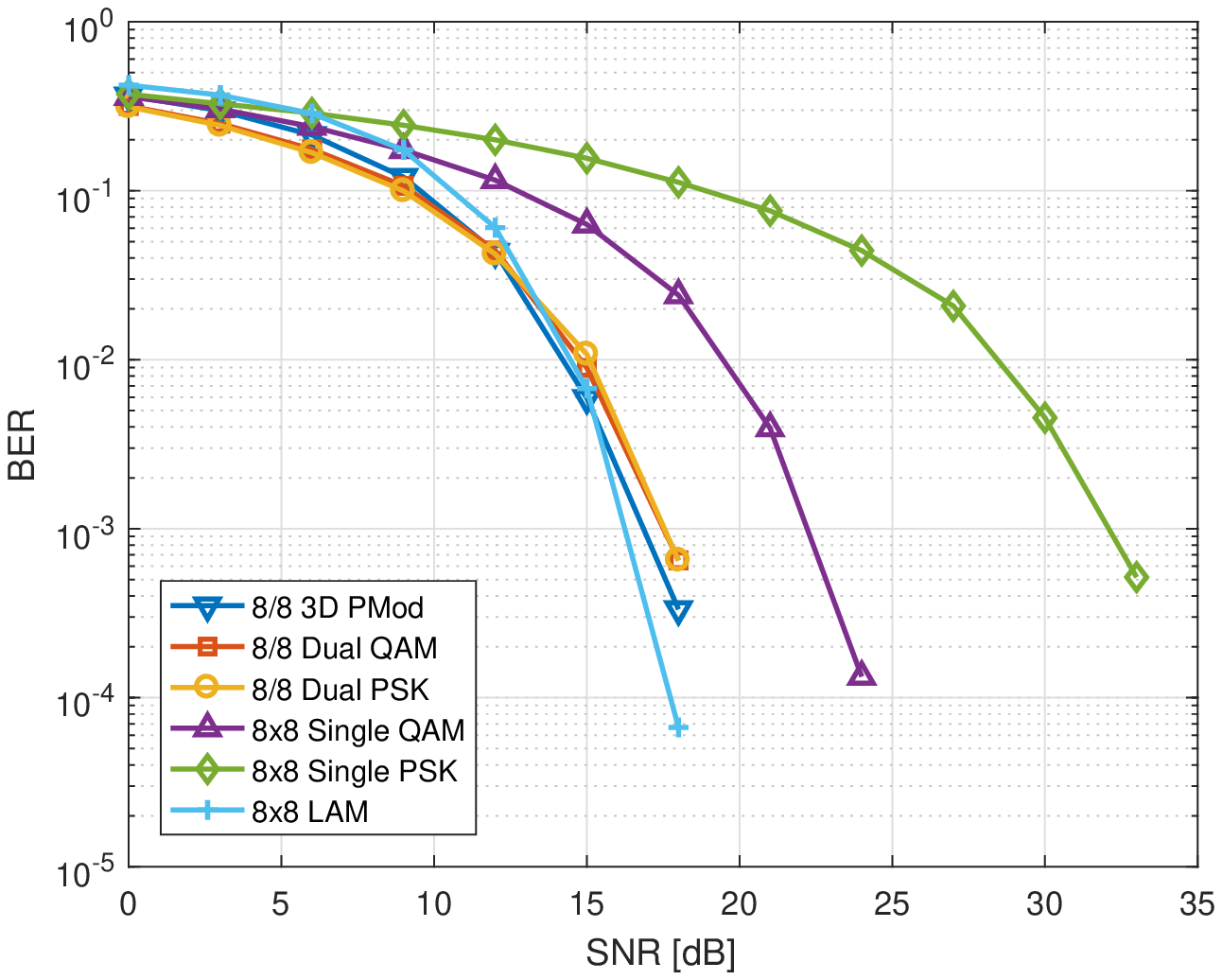}\label{fig:ber8x8}}\hfill
	\subfloat[][SE $7$ bps/Hz\\$L\times N=16\times 8$]{\includegraphics[width=0.45\linewidth,clip=true]{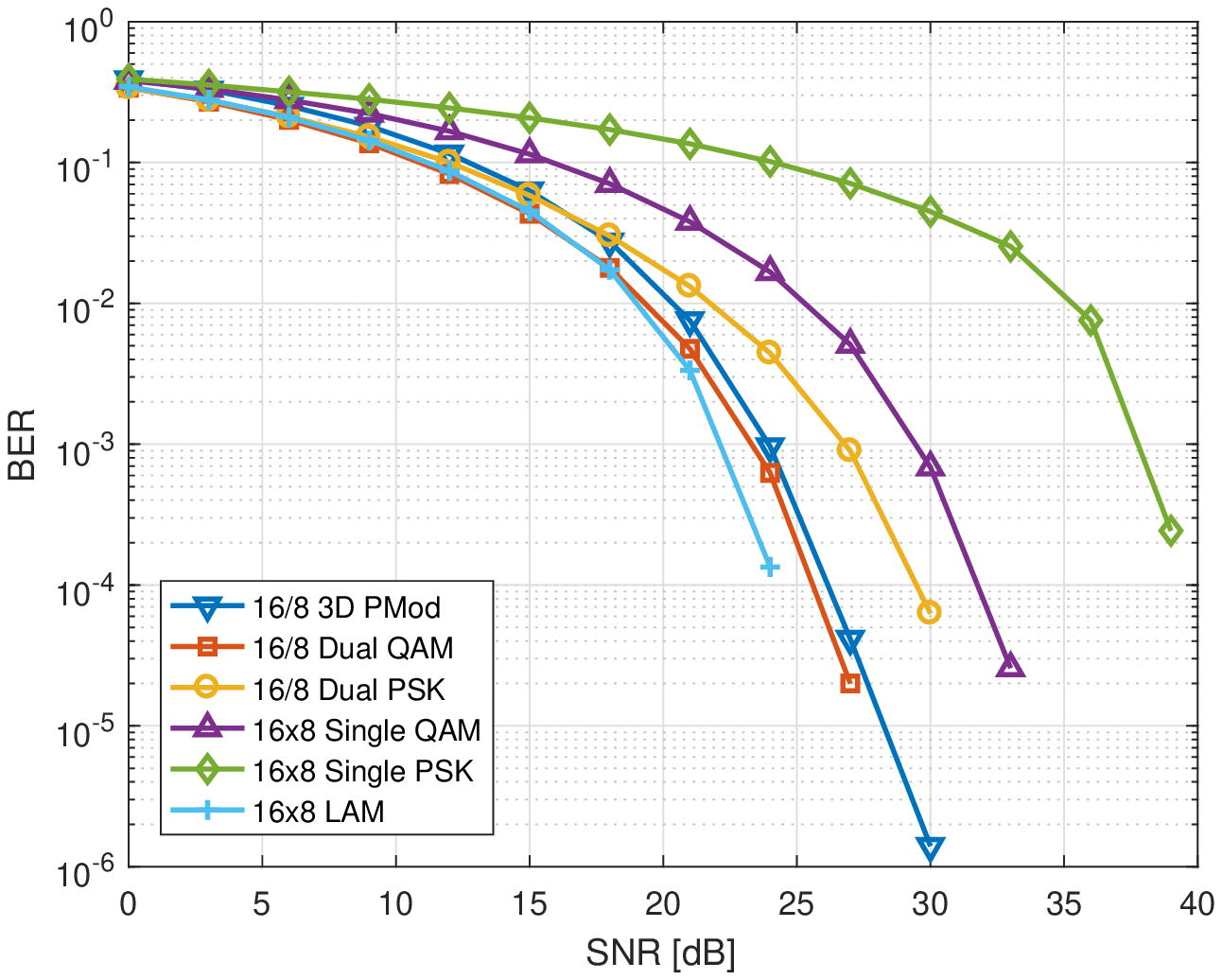}\label{fig:ber16x8}}\hfill   
	\caption{BER of 3D Polarized Modulation compared with other conventional schemes.}
	\label{fig:ber_3DPMod}
\end{figure}

Fig. \ref{fig:ber_3DPMod} illustrates the BER of 3D PMod compared with the aforementioned schemes for different constellation sizes. As we analyzed previously in terms of minimum distance, 3D PMod outperforms the other conventional schemes except LAM. Compared with LAM, 3D PMod achieves a similar performance, but with a higher degree of flexibility.


We also analyze the degradation of 2D PMod in front of 3D PMod. As mentioned above, 2D PMod obtains higher BER for $N>4$, compared with the optimal case of 3D PMod for the same spectral efficiency. 2D PMod is described in \cite{Henarejos2015a} and consists in transmitting a PSK/QAM mapped symbol activating horizontal or vertical polarization depending on the input source. In the previous section, we exposed that, in terms of minimum distance, 2D PMod achieves the higher minimum distance when a BPSK ($N=2$) or QPSK ($N=4$) is used as the symbol constellation, compared with other solutions.

\begin{figure}[!ht]
	\centering
	\subfloat[][SE $4$ bps/Hz]{\includegraphics[width=0.45\linewidth,clip=true]{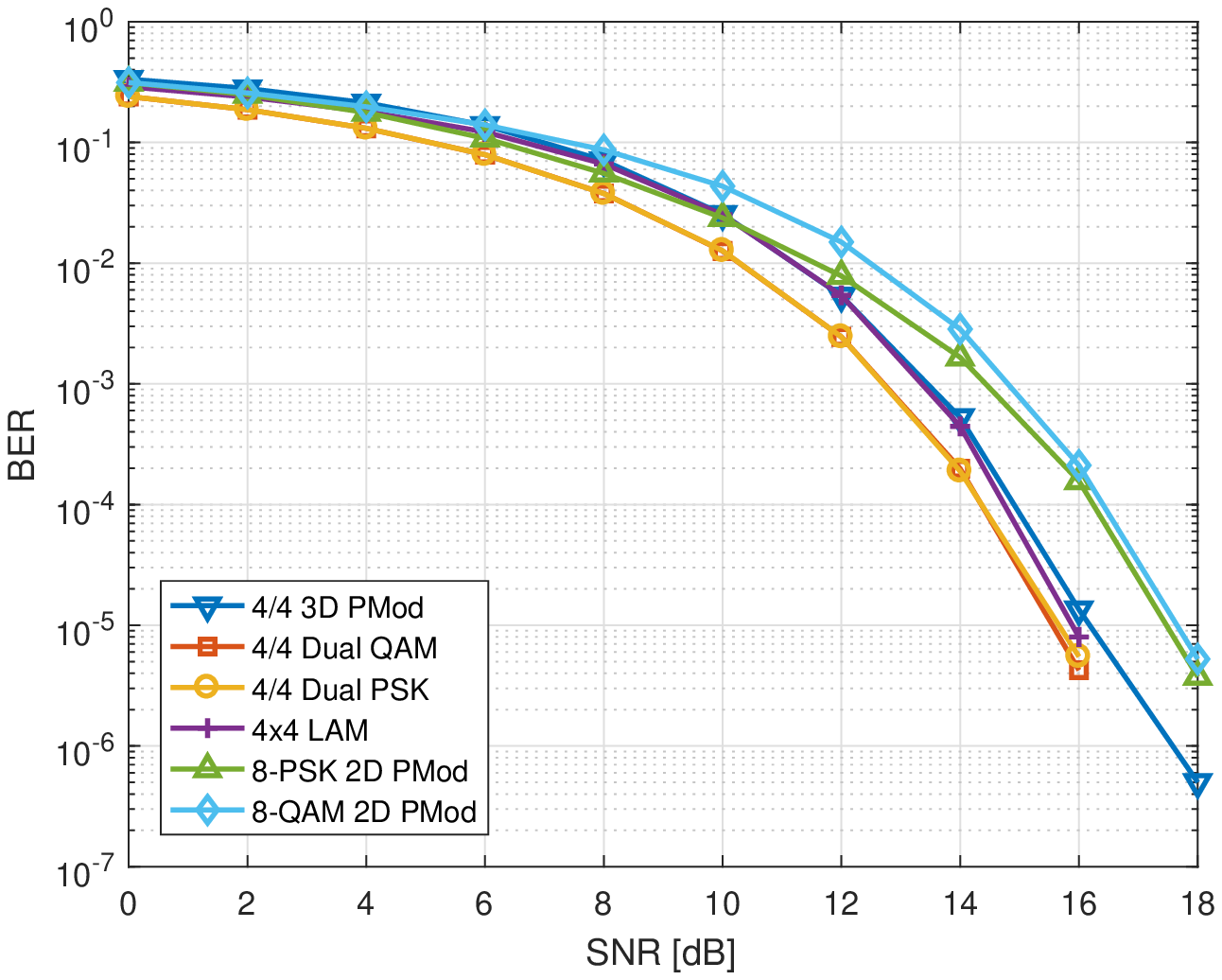}\label{fig:ber2D8}}\hfill
	\subfloat[][SE $5$ bps/Hz]{\includegraphics[width=0.45\linewidth,clip=true]{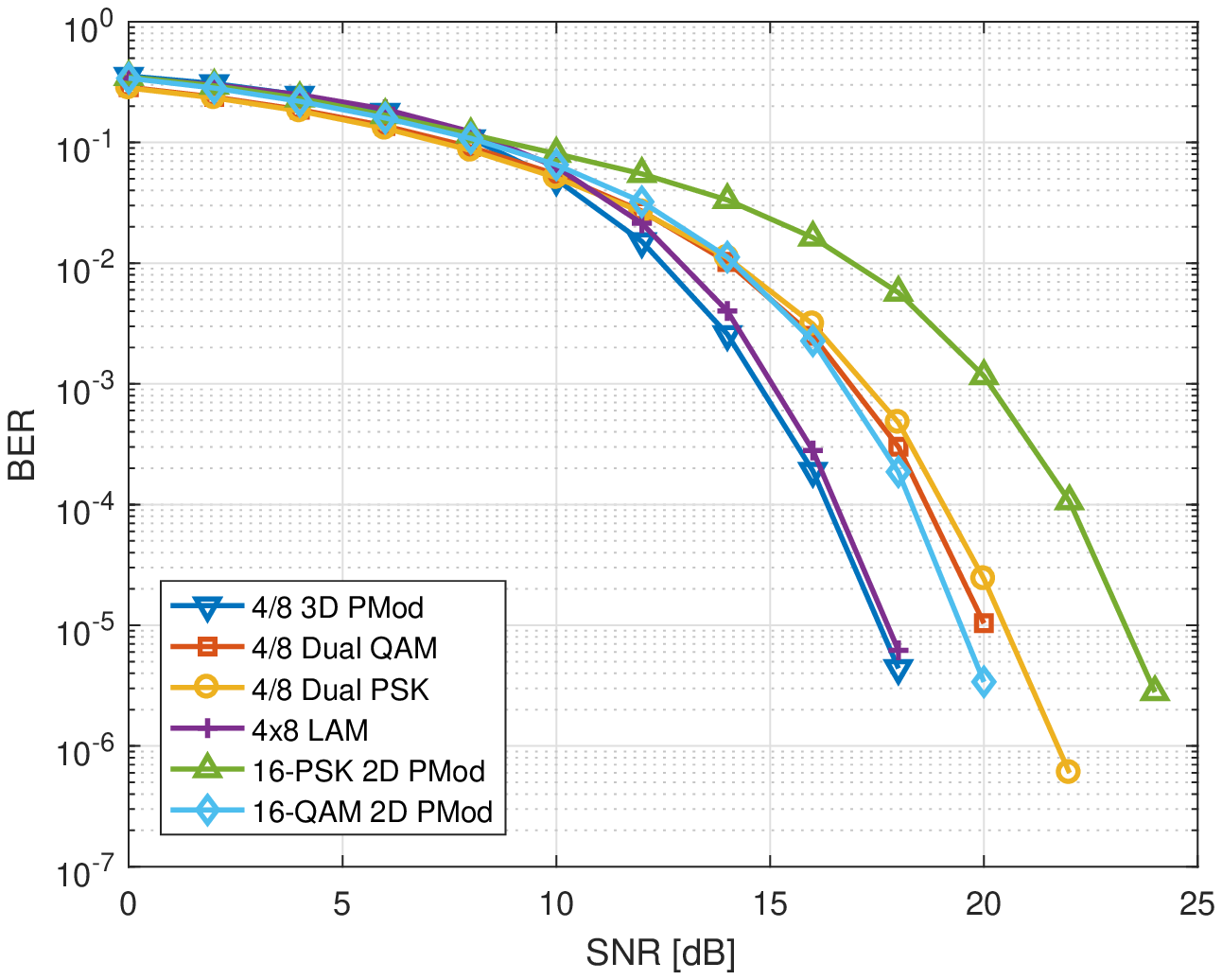}\label{fig:ber2D16}}\hfill
	\subfloat[][SE $6$ bps/Hz]{\includegraphics[width=0.45\linewidth,clip=true]{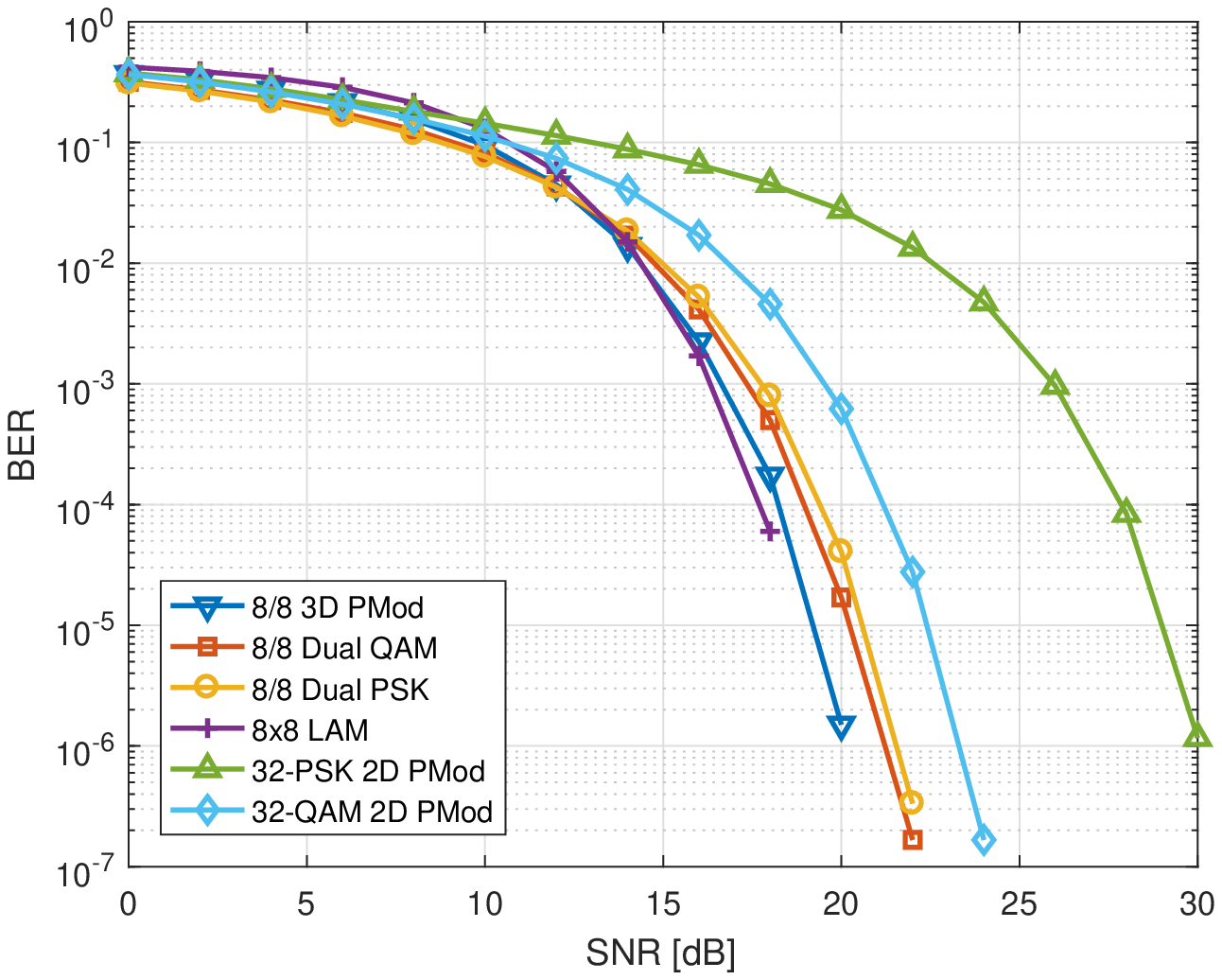}\label{fig:ber2D32}}\hfill
	\subfloat[][SE $7$ bps/Hz]{\includegraphics[width=0.45\linewidth,clip=true]{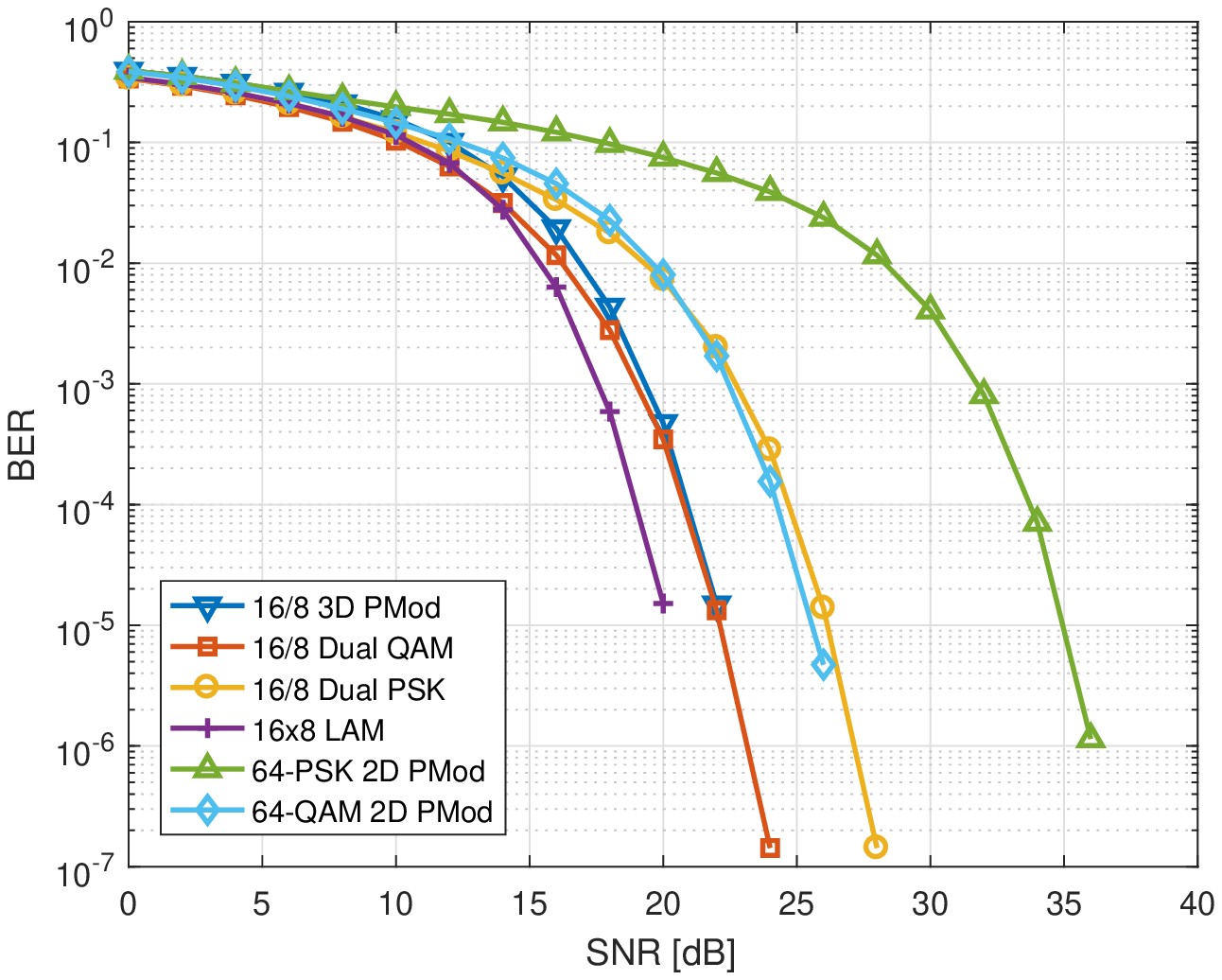}\label{fig:ber2D64}}\hfill
	\caption{BER of 2D PMod and 3D PMod with optimal mode for different spectral efficiencies.}
	\label{fig:ber_2DPMod}
\end{figure}

Fig. \ref{fig:ber_2DPMod} depicts the BER of 2D PMod and 3D PMod with optimal mode. As expected, 2D PMod performance is reduced notably when the spectral efficiency is increased. Moreover, since QAM has higher minimum distance than PSK, the performance of 2D-QAM PMod is higher than 2D-PSK PMod, though it is lower when it is compared with 3D PMod. The BER is also compared with Dual-QAM, Dual-PSK and LAM schemes.


\subsection{PDL and XPD}
In the previous sections, we compared the proposed schemes with other existing schemes considering an AWGN channel, i.e., $\mathbf{H}=\mathbf{I}_2$. In this section, we introduce the Polarization Dependent Loss (PDL) and Cross-Polarization Discrimination (XPD) effects. The former is produced when there is an imbalance of the power of each orthogonal polarization. The latter occurs when the signal from a polarization is coupled to the other. 

Although we assume that the receiver is able to estimate and equalize the channel, some errors and imperfections are still residual, affecting the performance of the system. Note that, in ideal conditions, the channel can be equalized to suppress its contribution but, in practice, this is not a realistic assumption and, thus, we model the channel as follows
\begin{equation}
\begin{split}
\mathbf{H}&=\mathbf{H}^{\textrm{XPD}}\mathbf{H}^{\textrm{PDL}}=\begin{pmatrix}1&\sqrt{(\zeta\psi)^{-1}}\\\sqrt{\zeta^{-1}}&\sqrt{\psi^{-1}}\end{pmatrix},
\end{split}
\label{eq:chanmod}
\end{equation}
where $\mathbf{H}^{\textrm{XPD}}=\begin{pmatrix}1&\sqrt{\zeta^{-1}}\\\sqrt{\zeta^{-1}}&1\end{pmatrix}$ and $0<\zeta\leq 1$ model the XPD; $\mathbf{H}^{\textrm{PDL}}=\begin{pmatrix}1&0\\0&\sqrt{\psi^{-1}}\end{pmatrix}$ and $0<\psi\leq 1$ model the PDL. For instance, if $\zeta\rightarrow 0$, there is no cross polarization, and if $\psi=1$, there is no PDL.

We remark that the model described in \eqref{eq:chanmod} contains all contributions that affect to XPD and PDL. For instance, although XPD models the discrimination between polarizations in the reflectors of satellites, it also contains the XPD degradation caused by rain drops \cite{Maitra2009} and other phenomena. Similarly, PDL not only contains the depolarization at the transmitter side but also the depolarization caused by other atmospheric aspects, such as Farady effect \cite{Lawrence2015}.

\begin{figure}[!ht]
	\centering
	\subfloat[][$\textrm{SNR}=9$dB, $\textrm{XPD}=4$dB]{\includegraphics[width=0.45\linewidth,clip=true]{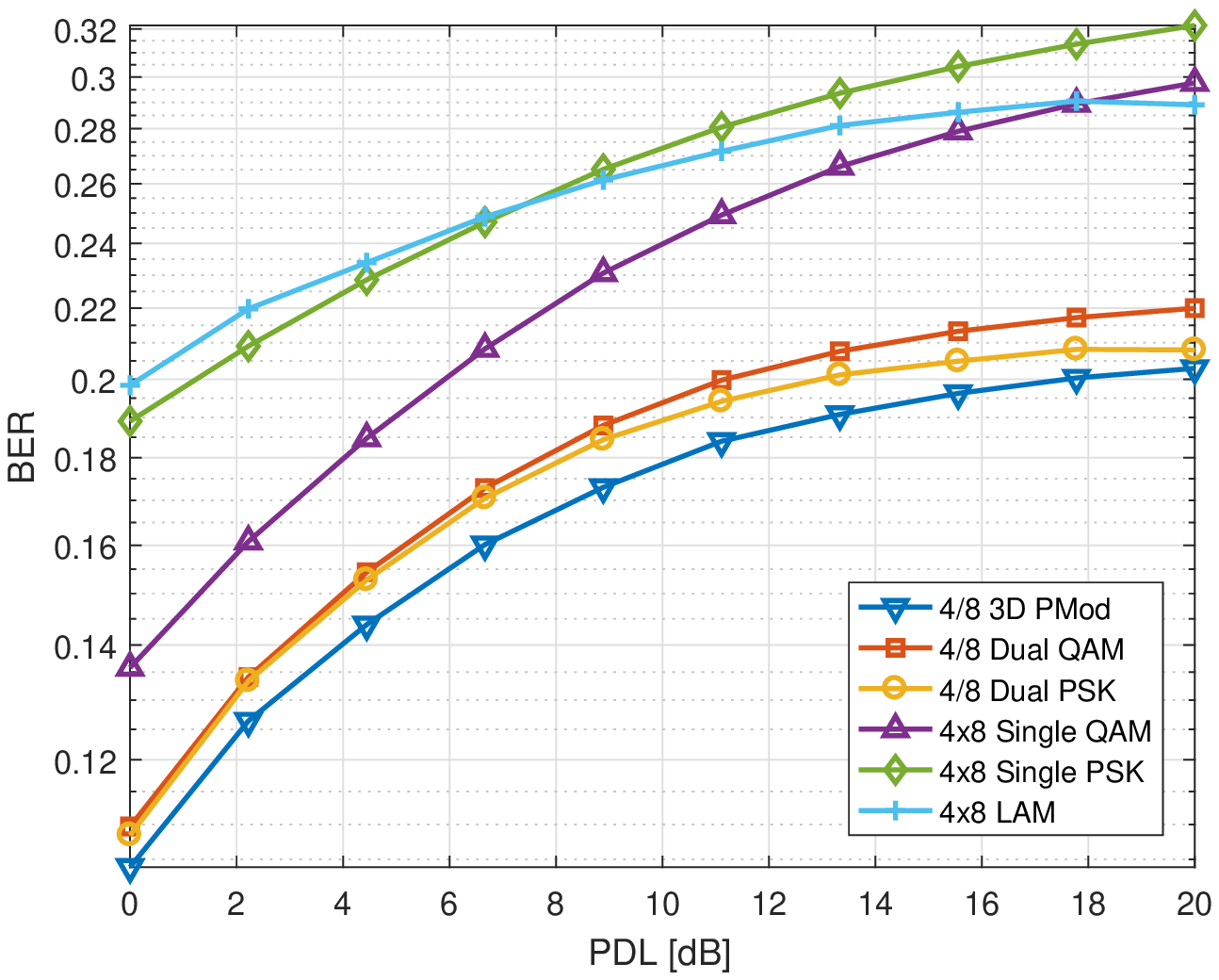}\label{fig:ber48_0904_pdl}}\hfill
	\subfloat[][$\textrm{SNR}=9$dB, $\textrm{XPD}=9$dB]{\includegraphics[width=0.45\linewidth,clip=true]{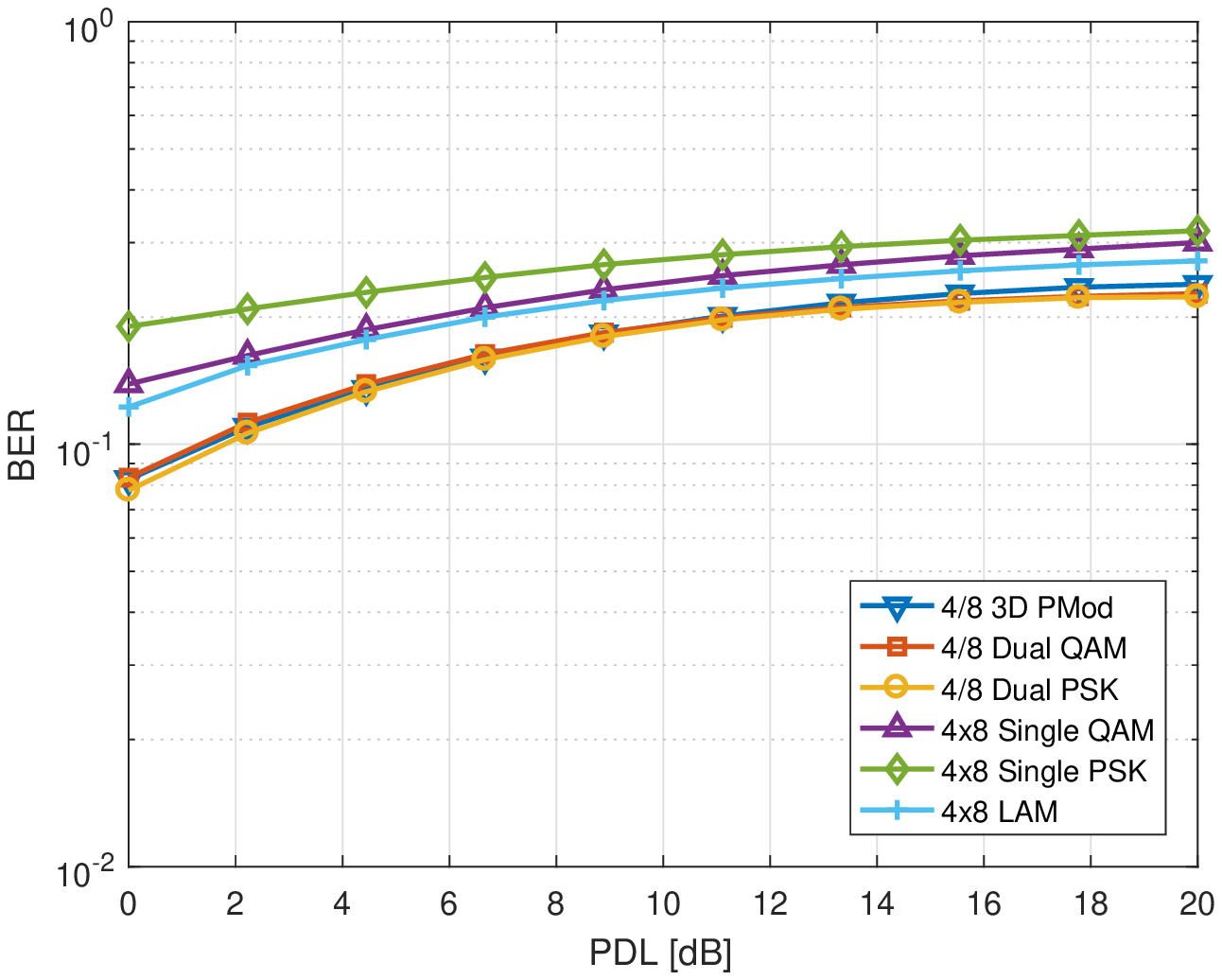}\label{fig:ber48_0909_pdl}}\hfill
	\subfloat[][$\textrm{SNR}=18$dB, $\textrm{XPD}=9$dB]{\includegraphics[width=0.45\linewidth,clip=true]{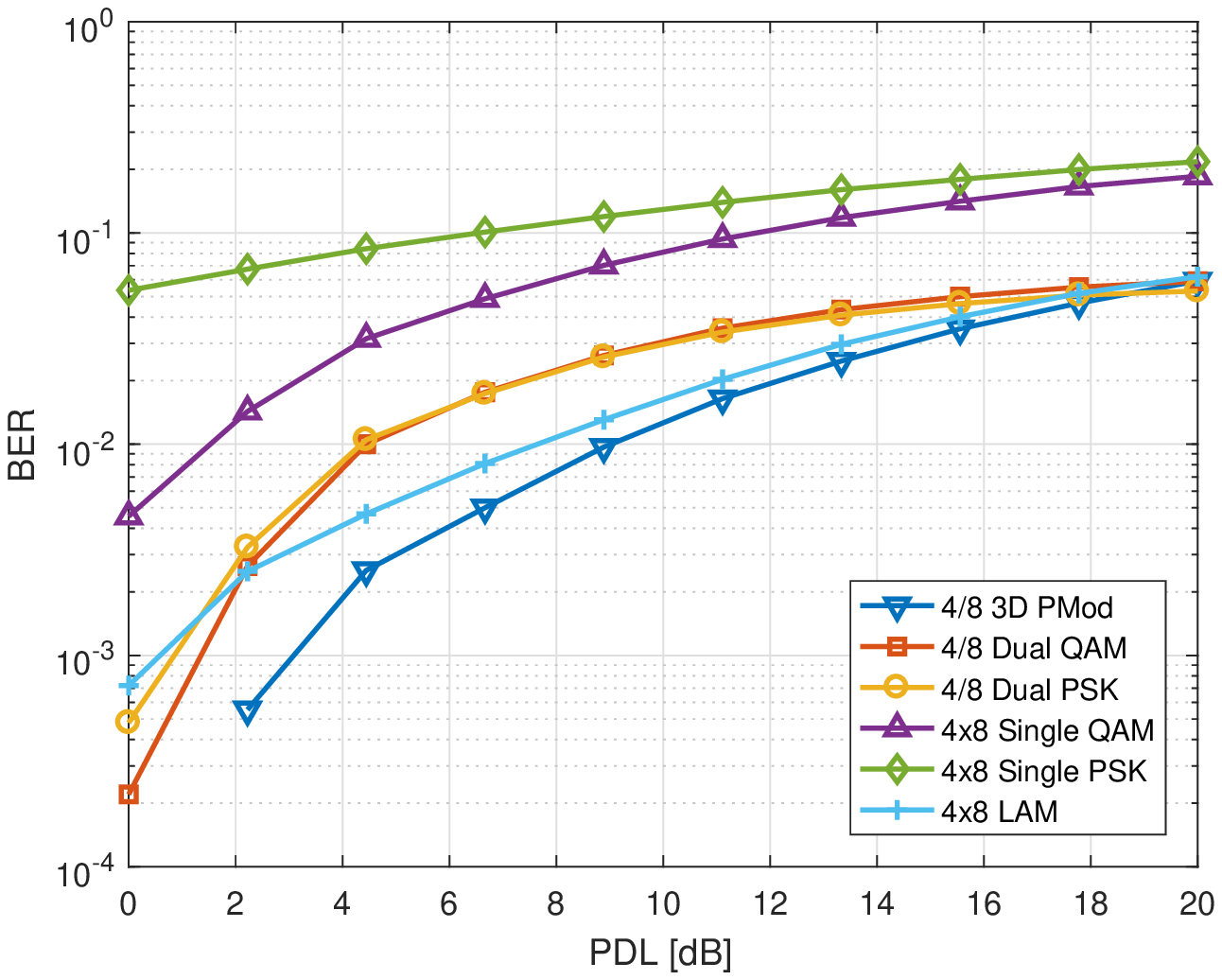}\label{fig:ber48_1809_pdl}}\hfill
	\subfloat[][$\textrm{SNR}=18$dB, $\textrm{XPD}=15$dB]{\includegraphics[width=0.45\linewidth,clip=true]{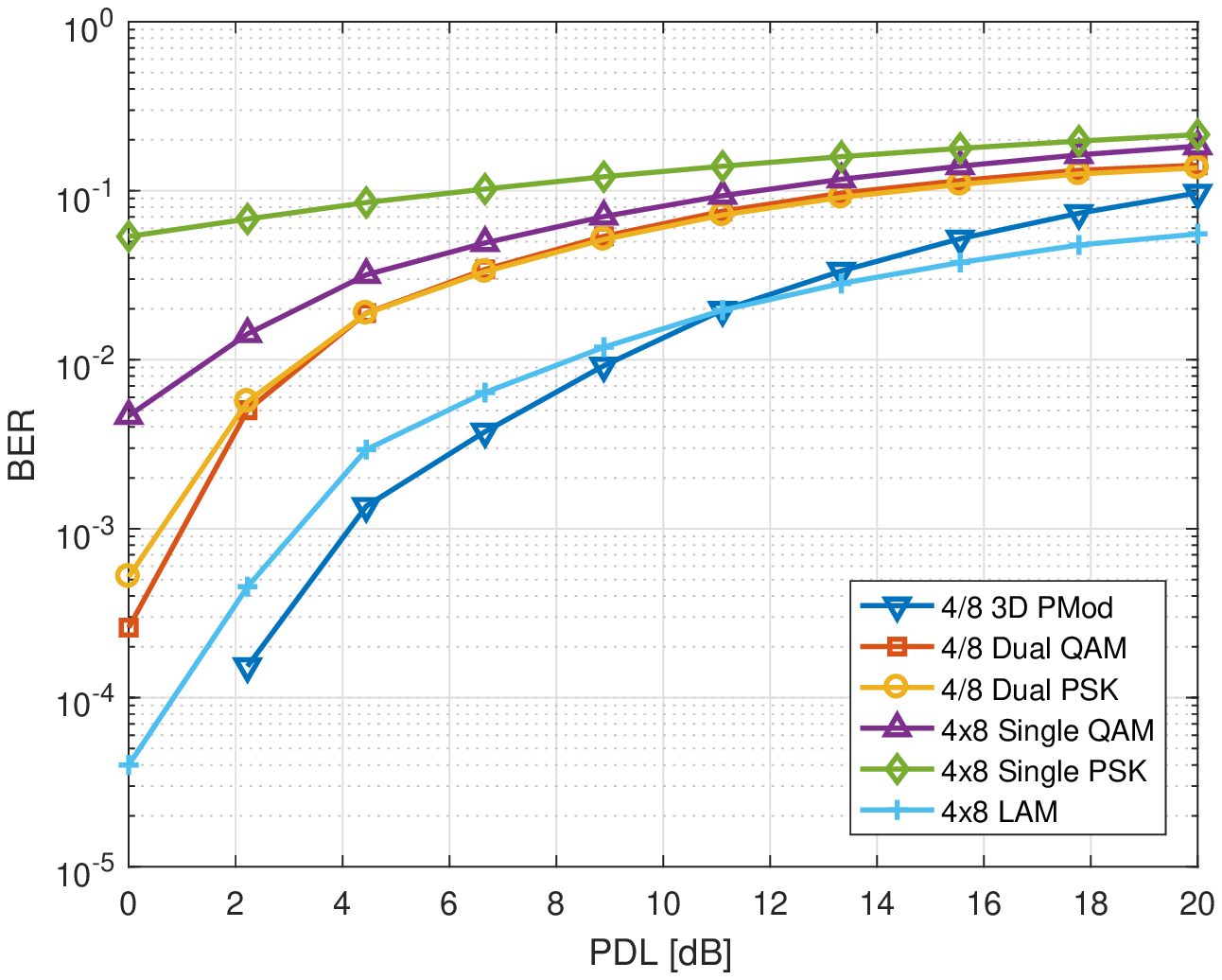}\label{fig:ber48_1820_pdl}}\hfill
	\caption{BER vs. PDL of different schemes conveying $5$ bps/Hz (3D PMod $4\times 8$), for different SNR and XPD values.}
	\label{fig:ber_PDL}
\end{figure}

Fig. \ref{fig:ber_PDL} depicts the BER depending on the PDL of different schemes conveying $5$ bps/Hz for different values of SNR and XPD. The first important aspect is that the performance of LAM, which is considered the optimal, is reduced drastically when XPD and PDL are considered. As expected, a we increase the PDL, the BER also increases. Nevertheless, as the figure shows, the proposed scheme is the most robust in front of PDL compared with other schemes. 

\begin{figure}[!ht]
	\centering
	\subfloat[][$\textrm{SNR}=9$dB, $\textrm{PDL}=0$dB]{\includegraphics[width=0.45\linewidth,clip=true]{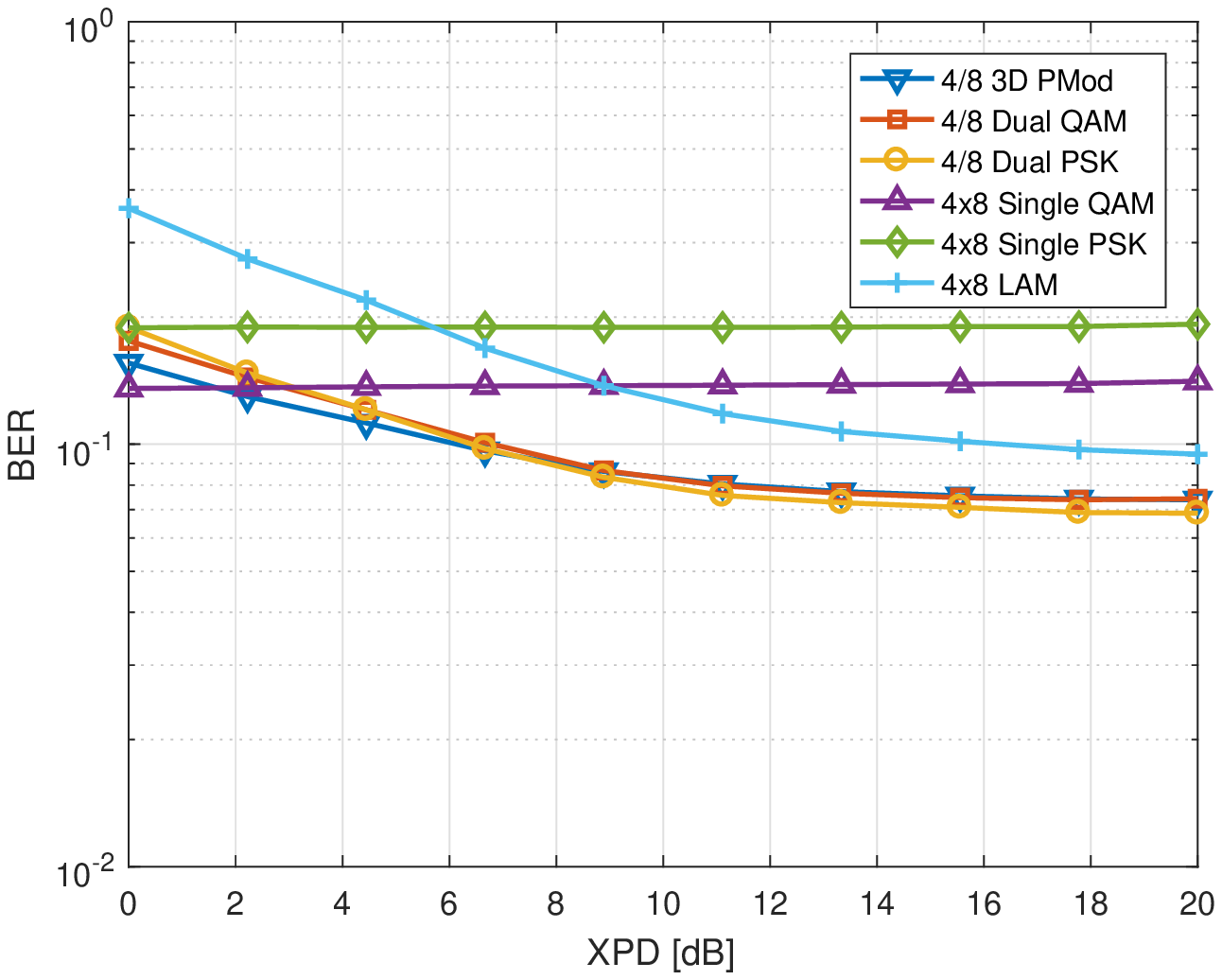}\label{fig:ber48_0900_xpd}}\hfill
	\subfloat[][$\textrm{SNR}=9$dB, $\textrm{PDL}=11$dB]{\includegraphics[width=0.45\linewidth,clip=true]{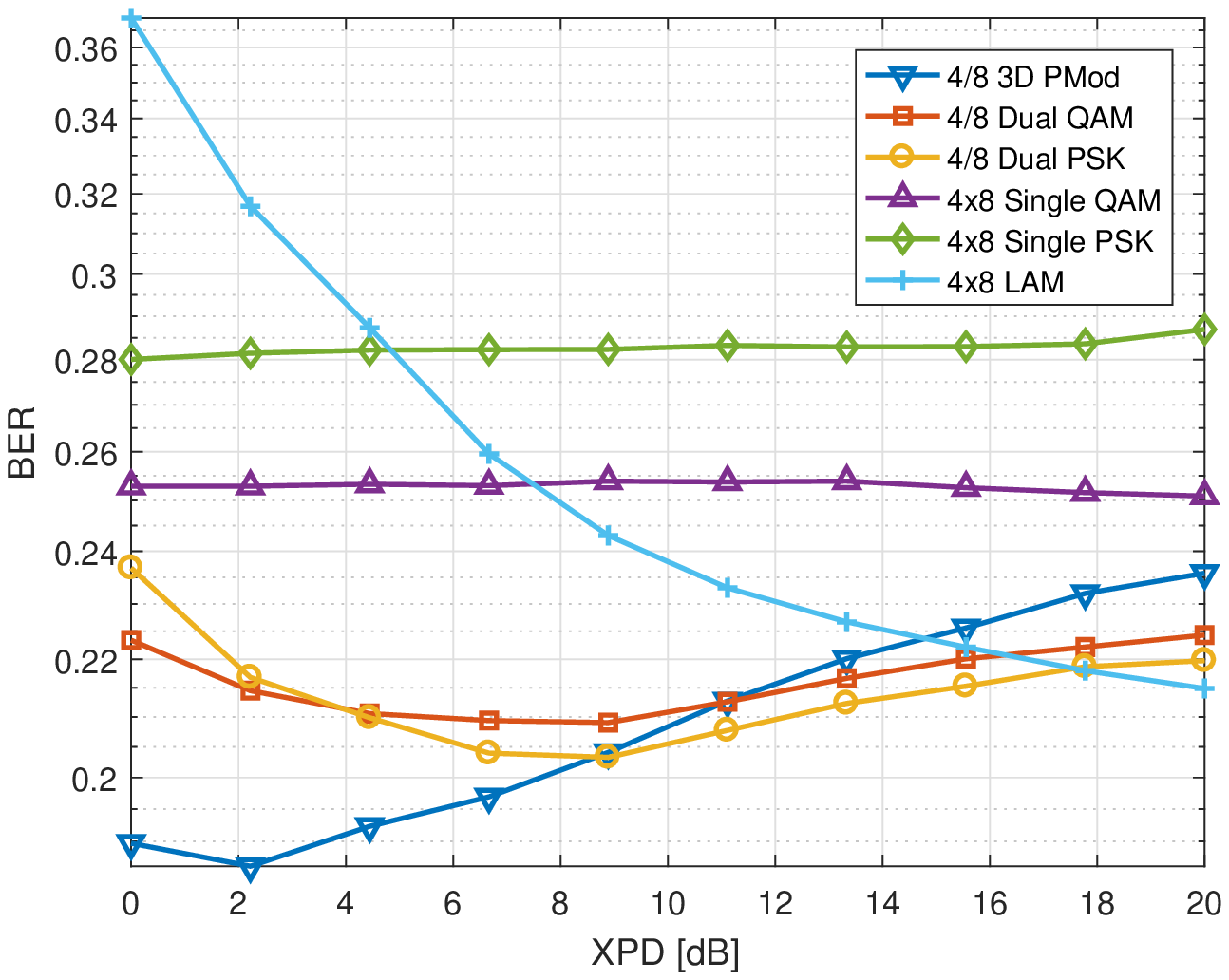}\label{fig:ber48_0911_xpd}}\hfill
	\subfloat[][$\textrm{SNR}=15$dB, $\textrm{PDL}=0$dB]{\includegraphics[width=0.45\linewidth,clip=true]{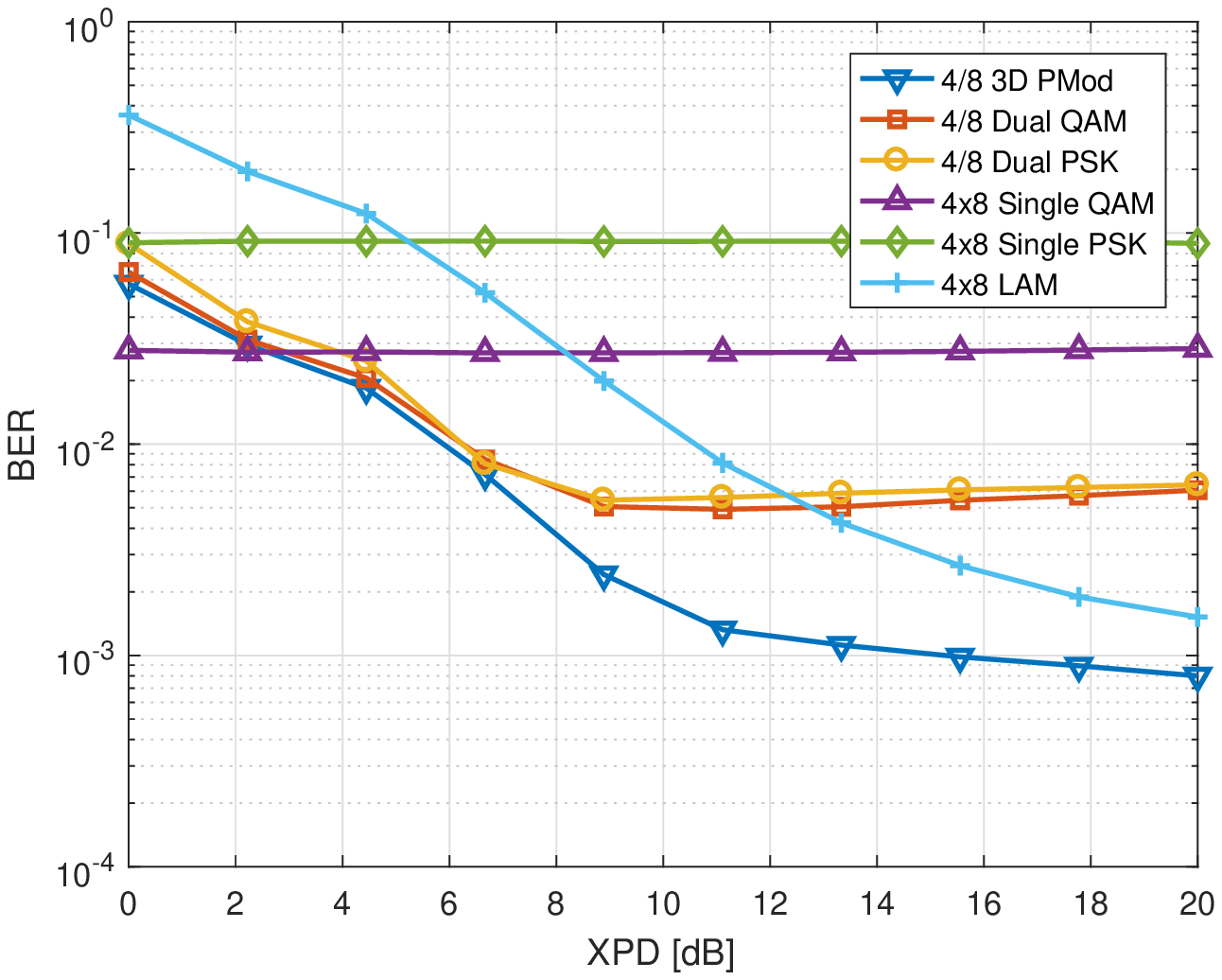}\label{fig:ber48_1500_xpd}}\hfill
	\subfloat[][$\textrm{SNR}=15$dB, $\textrm{PDL}=11$dB]{\includegraphics[width=0.45\linewidth,clip=true]{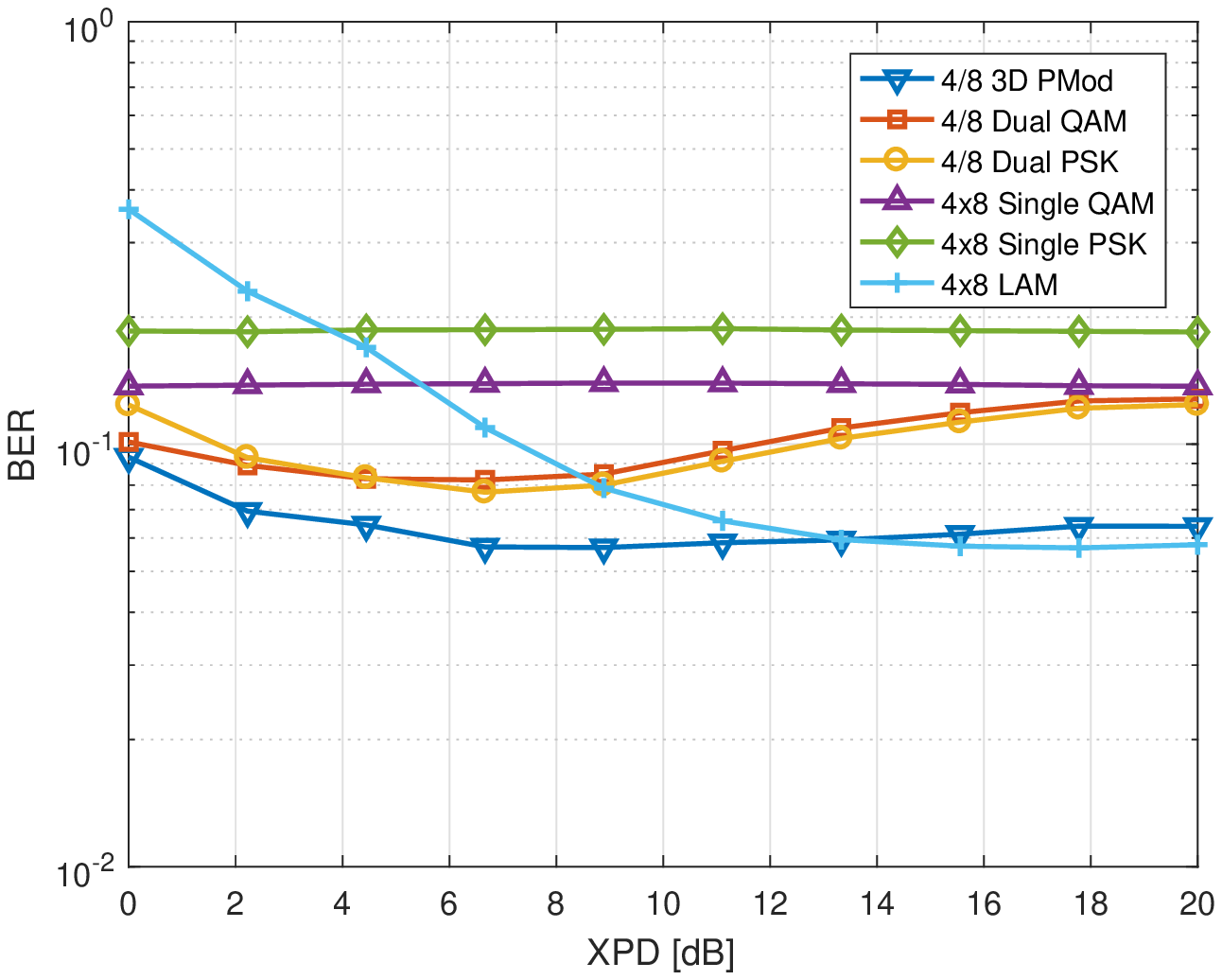}\label{fig:ber48_1511_xpd}}\hfill
	\caption{BER vs. XPD of different schemes conveying $5$ bps/Hz (3D PMod $4\times 8$), for different SNR and PDL values.}
	\label{fig:ber_XPD}
\end{figure}

Fig. \ref{fig:ber_XPD} illustrates the BER depending on the XPD of different schemes conveying $5$ bps/Hz for different values of SNR and PDL. On the one hand, examining all curves we can observe that for high values of XPD, the curves are asymptotic. This implies that, in absence of cross polarization, the predominant term is the SNR value, regardless the value of XPD. On the other hand, if PDL is predominant, the BER tends to increase for high XPD, which implies that PDL is the predominant term, as in Fig. \ref{fig:ber48_0911_xpd}. Moreover, we can appreciate that the 3D PMod also obtains the lowest BER compared with the other schemes and we can conclude that 3D PMod is more robust in front of PDL compared with the other schemes.

To summarize, we can observe the following advantages:
\begin{enumerate}
	\item Examining Fig. \ref{fig:minDist} we can observe that 3D PMod allows a finer adaptation when the SE is increased when it is compared with Dual or Single PSK/QAM schemes.
	\item 3D PMod achieves a performance near the optimal (LAM) but with much less computational complexity, for SE less than $8$ bps/Hz.
	\item Even with the presence of XPD and PDL, 3D PMod obtains higher performance compared with other schemes.
	\item 3D PMod allows flexible configurations to multiplex different streams with different Quality of Service. Thanks to this, the adaptation of allocated rate is smoother and finer.
	\item 3D PMod outperforms 2D PMod for SE above than $3$ bps/Hz.
\end{enumerate}

\section{Conclusions}
In this paper we present a new modulation based on the 3D constellation for polarization dimension. This modulation technique maps symbols from a sphere to the respective horizontal and vertical polarizations. This scheme is highly flexible since it allows to place an arbitrary number of symbols on the sphere and presents a low computational complexity. We describe the transmission scheme as well as two classes of receivers, depending on the performance and computational complexity trade-off. We study the analytical BER in terms of minimum distance and Union Bounds. We analyze the performance of the 3D PMod for different constellation sizes in terms of error rate. We compare the proposed classes of receivers and the performance of 3D PMod with other schemes such as Dual Polarization QAM multiplexing, Dual Polarization PSK multiplexing, Single Polarization QAM, Single Polarization PSK and LAM constellations. In this analysis, we consider realistic channel residual impairments such as PDL and XPD. Finally, we compare the proposed 3D PMod with conventional approaches of 2D PMod. From this, we emphasize that 3D PMod obtains the highest minimum distance for spectral efficiencies less than $8$ bps/Hz and in all cases, 3D PMod outperforms the other schemes, except for spectral efficiency of $4$ bps/Hz. Hence, we conclude that 3D PMod is an excellent option for medium and high modulation order transmissions.

\section*{Acknowledgement}
This work has received funding from the Spanish Ministry of Economy and Competitiveness (Ministerio de Economia y Competitividad) under project TERESA (TEC2017-90093-C3-1-R).

\printbibliography

@Article{Benedetto1992,
  author  = {Benedetto, S. and Poggiolini, P.},
  title   = {Theory of polarization shift keying modulation},
  journal = {IEEE Trans. Commun.},
  year    = {1992},
  volume  = {40},
  number  = {4},
  pages   = {708--721},
  doi     = {10.1109/26.141426},
}

@Article{Henarejos2015a,
  author  = {Henarejos, P. and Perez-Neira, A. I.},
  title   = {Dual Polarized Modulation and Reception for Next Generation Mobile Satellite Communications},
  journal = {IEEE Trans. Commun.},
  year    = {2015},
  volume  = {63},
  number  = {10},
  pages   = {3803--3812},
  doi     = {10.1109/TCOMM.2015.2461221},
}

@InProceedings{Henarejos2013,
  author    = {Henarejos, Pol and Vazquez, Miguel Angel and Cocco, Giuseppe and Perez-Neira, A. I. and others},
  title     = {Forward link interference mitigation in mobile interactive satellite systems},
  booktitle = {Proc. AIAA International Communications Satellite Systems Conference (ICSSC)},
  year      = {2013},
  month     = oct,
}

@Article{Fu2016,
  author   = {Y. Fu and C. X. Wang and Y. Yuan and R. Mesleh and e. H. M. Aggoune and M. M. Alwakeel and H. Haas},
  title    = {{BER} Performance of Spatial Modulation Systems Under {3-D} V2{V} {MIMO} Channel Models},
  journal  = {IEEE Trans. Veh. Technol.},
  year     = {2016},
  volume   = {65},
  number   = {7},
  pages    = {5725--5730},
  month    = jul,
  doi      = {10.1109/TVT.2015.2461638},
  issn     = {0018-9545},
}

@InProceedings{Henarejos2015,
  author    = {P. Henarejos and A. I. Perez-Neira},
  title     = {Dual Polarized Modulation and receivers for mobile communications in urban areas},
  booktitle = {Proc. IEEE 16th Int. Workshop Signal Processing Advances in Wireless Communications (SPAWC)},
  year      = {2015},
  pages     = {51--55},
  month     = jun,
  doi       = {10.1109/SPAWC.2015.7226998},
  issn      = {1948-3244},
}

@Misc{Proakis,
  author    = {Proakis, John G},
  title     = {Digital communications, 1995},
  publisher = {McGraw-Hill, New York},
}

@InProceedings{Wei2013,
  author       = {Wei, Dong and Feng, Chunyan and Guo, Caili},
  title        = {An optimal pre-compensation based joint polarization-amplitude-phase modulation scheme for the power amplifier energy efficiency improvement},
  booktitle    = {Communications (ICC), 2013 IEEE International Conference on},
  year         = {2013},
  pages        = {4137--4142},
  organization = {IEEE},
}

@Article{Arend2016,
  author    = {Arend, Lionel and Sperber, Ray and Marso, Michel and Krause, Jens},
  title     = {Implementing polarization shift keying over satellite--system design and measurement results},
  journal   = {Int. J. Satell. Commun. Networking},
  year      = {2016},
  volume    = {34},
  number    = {2},
  pages     = {211--229},
  publisher = {Wiley Online Library},
}

@Misc{Sloane,
  author = {N J A Sloane and R H Hardin and W D Smith and others},
  title  = {Tables of Spherical Codes},
  url    = {NeilSloane.com/packings},
}

@Article{Arapoglou2011,
  author   = {P. D. Arapoglou and P. Burzigotti and M. Bertinelli and A. Bolea Alamanac and R. De Gaudenzi},
  title    = {To {MIMO} or Not To {MIMO} in {Mobile Satellite Broadcasting Systems}},
  journal  = {IEEE Trans. Wireless Commun.},
  year     = {2011},
  volume   = {10},
  number   = {9},
  pages    = {2807--2811},
  month    = sep,
  issn     = {1536-1276},
  doi      = {10.1109/TWC.2011.071411.101599},
}

@InProceedings{Arapoglou2010,
  author    = {P. D. Arapoglou and P. Burzigotti and A. B. Alamanac and R. De Gaudenzi},
  title     = {Capacity potential of mobile satellite broadcasting systems employing dual polarization per beam},
  booktitle = {Proc. 5th Advanced Satellite Multimedia Systems Conf. and the 11th Signal Processing for Space Communications Workshop},
  year      = {2010},
  pages     = {213--220},
  month     = sep,
  doi       = {10.1109/ASMS-SPSC.2010.5586911},
  issn      = {2329-7093},
}

@InProceedings{Henarejos2016,
  author    = {P. Henarejos and A. I. Perez-Neira and N. Mazzali and C. Mosquera},
  title     = {Advanced signal processing techniques for fixed and mobile satellite communications},
  booktitle = {Proc. 8th Advanced Satellite Multimedia Systems Conf. and the 14th Signal Processing for Space Communications Workshop (ASMS/SPSC)},
  year      = {2016},
  pages     = {1--8},
  month     = sep,
  doi       = {10.1109/ASMS-SPSC.2016.7601468},
}

@Article{Renzo2014,
  author   = {M. Di Renzo and H. Haas and A. Ghrayeb and S. Sugiura and L. Hanzo},
  title    = {Spatial Modulation for Generalized {MIMO}: Challenges, Opportunities, and Implementation},
  journal  = {Proc. IEEE},
  year     = {2014},
  volume   = {102},
  number   = {1},
  pages    = {56--103},
  month    = jan,
  issn     = {0018-9219},
  doi      = {10.1109/JPROC.2013.2287851},
}

@InProceedings{Hardin1995,
  author       = {Hardin, RH and Sloane, NJA},
  title        = {Codes (spherical) and designs (experimental)},
  booktitle    = {Proceedings of Symposia in Applied Mathematics},
  year         = {1995},
  volume       = {50},
  pages        = {179--206},
  organization = {AMERICAN MATHEMATICAL SOCIETY},
}

@Book{Conway2013,
  title     = {Sphere packings, lattices and groups},
  publisher = {Springer Science \& Business Media},
  year      = {2013},
  author    = {Conway, John Horton and Sloane, Neil James Alexander},
  volume    = {290},
}

@Article{Sloane1998,
  author  = {Sloane, NJA},
  title   = {The sphere packing problem},
  journal = {Dimensions},
  year    = {1998},
  volume  = {4},
  number  = {8},
}

@Article{Renzo2012,
  author   = {M. Di Renzo and H. Haas},
  title    = {Bit Error Probability of {SM}-{MIMO} Over Generalized Fading Channels},
  journal  = {IEEE Trans. Veh. Technol.},
  year     = {2012},
  volume   = {61},
  number   = {3},
  pages    = {1124--1144},
  month    = mar,
  issn     = {0018-9545},
  doi      = {10.1109/TVT.2012.2186158},
}

@Article{Tammes1930,
  author    = {Tammes, Pieter Merkus Lambertus},
  title     = {On the origin of number and arrangement of the places of exit on the surface of pollen-grains},
  journal   = {Recueil des travaux botaniques neerlandais},
  year      = {1930},
  volume    = {27},
  number    = {1},
  pages     = {1--84},
  publisher = {Koninklijke Nederlandse Botanische Vereniging},
}

@Article{Kottwitz1991,
  author    = {Kottwitz, DA},
  title     = {The densest packing of equal circles on a sphere},
  journal   = {Acta Crystallographica Section A: Foundations of Crystallography},
  year      = {1991},
  volume    = {47},
  number    = {3},
  pages     = {158--165},
  publisher = {International Union of Crystallography},
}

@InProceedings{Nomura2011,
  author       = {Nomura, Hiroshi and Takahashi, Masanori and Kyoh, Suigen},
  title        = {Physical conversion of {S}tokes parameters which are multiplied by a general {M}ueller matrix into {J}ones vectors applicable to the lithographic calculation},
  booktitle    = {SPIE Advanced Lithography},
  year         = {2011},
  pages        = {79731O--79731O},
  organization = {International Society for Optics and Photonics},
}

@InProceedings{Mazzali2016,
  author       = {Mazzali, Nicolo and Kayhan, Farbod and Mysore, R Bhavani Shankar},
  title        = {Four-dimensional constellations for dual-polarized satellite communications},
  booktitle    = {Communications (ICC), 2016 IEEE International Conference on},
  year         = {2016},
  pages        = {1--6},
  organization = {IEEE},
}

@Article{Wei2013a,
  author   = {D. Wei and C. Feng and C. Guo and L. Fangfang},
  title    = {A Power Amplifier Energy Efficient Polarization Modulation Scheme Based on the Optimal Pre-Compensation},
  journal  = {IEEE Commun. Lett.},
  year     = {2013},
  volume   = {17},
  number   = {3},
  pages    = {513--516},
  month    = mar,
  issn     = {1089-7798},
  doi      = {10.1109/LCOMM.2013.012313.122639},
}

@InProceedings{Wei2012,
  author    = {D. Wei and C. Feng and C. Guo},
  title     = {A Polarization-Amplitude-Phase Modulation scheme for the Power Amplifier energy efficiency enhancement},
  booktitle = {Proc. 15th Int. Symp. Wireless Personal Multimedia Communications},
  year      = {2012},
  pages     = {369--373},
  month     = sep,
  issn      = {1347-6890},
}

@InProceedings{Wei2016,
  author    = {Dong Wei and Lili Liang and Meng Zhang and Rong Qiao and Weiqing Huang},
  title     = {A Polarization state Modulation based Physical Layer Security scheme for Wireless Communications},
  booktitle = {Proc. MILCOM 2016 - 2016 IEEE Military Communications Conf},
  year      = {2016},
  pages     = {1195--1201},
  month     = nov,
  doi       = {10.1109/MILCOM.2016.7795493},
}

@Article{Stokes1852,
  author    = {Stokes, George Gabriel},
  title     = {On the change of refrangibility of light},
  journal   = {Philosophical Transactions of the Royal Society of London},
  year      = {1852},
  volume    = {142},
  pages     = {463--562},
  publisher = {JSTOR},
}

@Book{Poincare1892,
  title     = {Theorie mathematique de la lumiere},
  publisher = {Gauthier Villars},
  year      = {1892},
  author    = {Poincare, Henri},
  volume    = {2},
}

@Article{Benedetto1994,
  author   = {S. Benedetto and P. T. Poggiolini},
  title    = {Multilevel polarization shift keying: optimum receiver structure and performance evaluation},
  journal  = {IEEE Trans. Commun.},
  year     = {1994},
  volume   = {42},
  number   = {234},
  pages    = {1174--1186},
  month    = feb,
  issn     = {0090-6778},
  doi      = {10.1109/TCOMM.1994.580226},
}

@Article{Cao2011,
  author    = {Cao, Bin and Zhang, Qin-Yu and Jin, Lin},
  title     = {Polarization division multiple access with polarization modulation for LOS wireless communications},
  journal   = {EURASIP Journal on wireless communications and networking},
  year      = {2011},
  volume    = {2011},
  number    = {1},
  pages     = {77},
  publisher = {Springer},
}

@Article{Zafari2017,
  author   = {G. Zafari and M. Koca and H. Sari},
  title    = {Dual-Polarized Spatial Modulation Over Correlated Fading Channels},
  journal  = {IEEE Trans. Commun.},
  year     = {2017},
  volume   = {65},
  number   = {3},
  pages    = {1336--1352},
  month    = mar,
  issn     = {0090-6778},
  doi      = {10.1109/TCOMM.2016.2643664},
}

@Article{Guo2016,
  author   = {S. Guo and H. Zhang and S. Jin and P. Zhang},
  title    = {Spatial Modulation via {3-D} Mapping},
  journal  = {IEEE Commun. Lett.},
  year     = {2016},
  volume   = {20},
  number   = {6},
  pages    = {1096--1099},
  month    = jun,
  issn     = {1089-7798},
  doi      = {10.1109/LCOMM.2016.2556685},
}

@Article{Gu2017,
  author   = {J. F. Gu and K. Wu},
  title    = {Quaternion Modulation for Dual-Polarized Antennas},
  journal  = {IEEE Commun. Lett.},
  year     = {2017},
  volume   = {21},
  number   = {2},
  pages    = {286--289},
  month    = feb,
  issn     = {1089-7798},
  doi      = {10.1109/LCOMM.2016.2614512},
}

@Article{Zhang2017a,
  author   = {J. Zhang and Y. Wang and J. Zhang and L. Ding},
  title    = {Polarization Shift Keying (PolarSK): System Scheme and Performance Analysis},
  journal  = {IEEE Trans. Veh. Technol.},
  year     = {2017},
  volume   = {PP},
  number   = {99},
  pages    = {1},
  issn     = {0018-9545},
  doi      = {10.1109/TVT.2017.2755401},
}

@InProceedings{Nie2014,
  author    = {Y. Nie and C. Feng and C. Guo and F. Liu},
  title     = {Common phase error cancellation scheme with DPOLSK in {OFDM} system},
  booktitle = {Proc. 21st Int. Conf. Telecommunications (ICT)},
  year      = {2014},
  pages     = {129--133},
  month     = may,
  doi       = {10.1109/ICT.2014.6845094},
}

@Article{Henarejos2017,
  author   = {P. Henarejos and A. I. Perez-Neira},
  title    = {Capacity Analysis of Index Modulations Over Spatial, Polarization, and Frequency Dimensions},
  journal  = {IEEE Trans. Commun.},
  year     = {2017},
  volume   = {65},
  number   = {12},
  pages    = {5280--5292},
  month    = dec,
  issn     = {0090-6778},
  doi      = {10.1109/TCOMM.2017.2743166},
}

@Article{Maitra2009,
  author   = {A. Maitra and K. Chakravarty},
  title    = {Rain Depolarization Measurements on Low Margin Ku-Band Satellite Signal at a Tropical Location},
  journal  = {IEEE Antennas Wireless Propag. Lett.},
  year     = {2009},
  volume   = {8},
  pages    = {445--448},
  issn     = {1536-1225},
  doi      = {10.1109/LAWP.2009.2015969},
}

@Article{Lawrence2015,
  author   = {N. P. Lawrence and H. J. Hansen and D. Abbott},
  title    = {{3-D} Low Earth Orbit Vector Estimation of Faraday Rotation and Path Delay},
  journal  = {IEEE Access},
  year     = {2015},
  volume   = {3},
  pages    = {1684--1694},
  issn     = {2169-3536},
  doi      = {10.1109/ACCESS.2015.2479247},
}

@InProceedings{Elkawafi2017,
  author        = {S. Elkawafi and A. Younis and R. Mesleh and A. Abouda and A. Elbarsha and M. Elmusrati},
  title         = {Spatial Modulation and Spatial Multiplexing Capacity Analysis over {3D} mmWave Communications},
  booktitle     = {Proc. European Wireless 2017; 23th European Wireless Conf},
  year          = {2017},
  pages         = {1--6},
  month         = may,
  __markedentry = {[Pol:]},
}

@Article{Guo2017,
  author        = {C. Guo and F. Liu and S. Chen and C. Feng and Z. Zeng},
  title         = {Advances on Exploiting Polarization in Wireless Communications: Channels, Technologies, and Applications},
  journal       = {IEEE Communications Surveys Tutorials},
  year          = {2017},
  volume        = {19},
  number        = {1},
  pages         = {125--166},
  month         = fir,
  issn          = {1553-877X},
  __markedentry = {[Pol:6]},
  doi           = {10.1109/COMST.2016.2606639},
  keywords      = {channel capacity, cognitive radio, electromagnetic wave polarisation, energy conservation, telecommunication power management, wireless channels, channel capacity enhancement, cognitive radio networks, complex depolarization effect, electromagnetic wave polarization, energy-efficient communication, in-band full-duplex transmission, optical fiber, orthogonally polarized electric dipole, orthogonally polarized magnetic dipole, polarization channel, polarization domain, radar, satellite communication, wireless channel, wireless communication, Laser radar, Modulation, Optical polarization, Optical sensors, Tutorials, Wireless communication, Wireless communications, cognitive radio (CR) networks, energy-efficient communications, in-band full-duplex (IBFD) transmission, polarization channels, polarization technologies},
}

\begin{table}[!ht]
	\renewcommand{\arraystretch}{1.2}
	\centering
	\caption{Minimum distance for spectral efficiency $L_b+N_b=2$ bits}
	\begin{tabular}{c|c|c|c|c|c|c}
		$L\times N$ & 3D PMod & \multiline{Dual\\QAM} & \multiline{Dual\\PSK} & \multiline{Single\\QAM} & \multiline{Single\\PSK} & LAM \\\hline
		$2\times 2$ & $\mathbf{1.4142}$ & $1.4142$ & $1.4142$ & $1.4142$ & $1.4142$ & $1.4142$\\\hline
	\end{tabular}
	\label{tab:mindist2}
\end{table}

\begin{table}[!ht]
	\renewcommand{\arraystretch}{1.2}
	\centering
	\caption{Minimum distance for spectral efficiency $L_b+N_b=3$ bits}
	\begin{tabular}{c|c|c|c|c|c|c}
		$L\times N$ & 3D PMod & \multiline{Dual\\QAM} & \multiline{Dual\\PSK} & \multiline{Single\\QAM} & \multiline{Single\\PSK} & LAM \\\hline
		$2\times 4$ & $\mathbf{1.4142}$ & $1$ & $1$ & $0.8165$ & $0.7654$ & $1.4142$\\\hline
		$4\times 2$ & $1$ & $1$ & $1$ & $0.8165$ & $0.7654$ & $1.4142$\\\hline
	\end{tabular}
	\label{tab:mindist3}
\end{table}

\begin{table}[!ht]
	\renewcommand{\arraystretch}{1.2}
	\centering
	\caption{Minimum distance for spectral efficiency $L_b+N_b=4$ bits}
	\begin{tabular}{c|c|c|c|c|c|c}
		$L\times N$ & 3D PMod & \multiline{Dual\\QAM} & \multiline{Dual\\PSK} & \multiline{Single\\QAM} & \multiline{Single\\PSK} & LAM \\\hline
		$2\times 8$ & $0.7654$ & $0.5774$ & $0.5412$ & $0.6325$ & $0.3902$ & $1$\\\hline
		$8\times 2$ & $0.6323$ & $0.5774$ & $0.5412$ & $0.6325$ & $0.3902$ & $1$\\\hline
		$4\times 4$ & $0.9194$ & $\mathbf{1}$ & $\mathbf{1}$ & $0.6325$ & $0.3902$ & $1$\\\hline
	\end{tabular}
	\label{tab:mindist4}
\end{table}

\begin{table}[!ht]
	\renewcommand{\arraystretch}{1.2}
	\centering
	\caption{Minimum distance for spectral efficiency $L_b+N_b=5$ bits}
	\begin{tabular}{c|c|c|c|c|c|c}
		$L\times N$ & 3D PMod & \multiline{Dual\\QAM} & \multiline{Dual\\PSK} & \multiline{Single\\QAM} & \multiline{Single\\PSK} & LAM \\\hline
		$2\times 16$ & $0.3902$ & $0.4472$ & $0.2759$ & $0.4472$ & $0.1960$ & $0.8165$\\\hline
		$16\times 2$ & $0.5039$ & $0.4472$ & $0.2759$ & $0.4472$ & $0.1960$ & $0.8165$\\\hline
		$4\times 8$ & $\mathbf{0.7654}$ & $0.5774$ & $0.5412$ & $0.4472$ & $0.1960$ & $0.8165$\\\hline
		$8\times 4$ & $0.6323$ & $0.5774$ & $0.5412$ & $0.4472$ & $0.1960$ & $0.8165$\\\hline
	\end{tabular}
	\label{tab:mindist5}
\end{table}

\begin{table}[!ht]
	\renewcommand{\arraystretch}{1.2}
	\centering
	\caption{Minimum distance for spectral efficiency $L_b+N_b=6$ bits}
	\begin{tabular}{c|c|c|c|c|c|c}
		$L\times N$ & 3D PMod & \multiline{Dual\\QAM} & \multiline{Dual\\PSK} & \multiline{Single\\QAM} & \multiline{Single\\PSK} & LAM \\\hline
		$2\times 32$ & $0.1960$ & $0.3162$ & $0.1386$ & $0.3086$ & $0.0981$ & $0.7559$\\\hline
		$32\times 2$ & $0.3318$ & $0.3162$ & $0.1386$ & $0.3086$ & $0.0981$ & $0.7559$\\\hline
		$4\times 16$ & $0.3902$ & $0.4472$ & $0.2759$ & $0.3086$ & $0.0981$ & $0.7559$\\\hline
		$16\times 4$ & $0.5039$ & $0.4472$ & $0.2759$ & $0.3086$ & $0.0981$ & $0.7559$\\\hline
		$8\times 8$ & $\mathbf{0.6323}$ & $0.5774$ & $0.5412$ & $0.3086$ & $0.0981$ & $0.7559$\\\hline
	\end{tabular}
	\label{tab:mindist6}
\end{table}

\begin{table}[!ht]
	\renewcommand{\arraystretch}{1.2}
	\centering
	\caption{Minimum distance for spectral efficiency $L_b+N_b=7$ bits}
	\begin{tabular}{c|c|c|c|c|c|c}
		$L\times N$ & 3D PMod & \multiline{Dual\\QAM} & \multiline{Dual\\PSK} & \multiline{Single\\QAM} & \multiline{Single\\PSK} & LAM \\\hline
		$2\times 64$ & $0.0981$ & $0.2182$ & $0.0694$ & $0.2209$ & $0.0491$ & $0.6324$\\\hline
		$64\times 2$ & $0.2615$ & $0.2182$ & $0.0694$ & $0.2209$ & $0.0491$ & $0.6324$\\\hline
		$4\times 32$ & $0.1960$ & $0.3162$ & $0.1386$ & $0.2209$ & $0.0491$ & $0.6324$\\\hline
		$32\times 4$ & $0.3318$ & $0.3162$ & $0.1386$ & $0.2209$ & $0.0491$ & $0.6324$\\\hline
		$8\times 16$ & $0.3902$ & $0.4472$ & $0.2759$ & $0.2209$ & $0.0491$ & $0.6324$\\\hline
		$16\times 8$ & $\mathbf{0.4627}$ & $0.4472$ & $0.2759$ & $0.2209$ & $0.0491$ & $0.6324$\\\hline
	\end{tabular}
	\label{tab:mindist7}
\end{table}

\begin{table}[!ht]
	\renewcommand{\arraystretch}{1.2}
	\centering
	\caption{Minimum distance for spectral efficiency $L_b+N_b=8$ bits}
	\begin{tabular}{c|c|c|c|c|c|c}
		$L\times N$ & 3D PMod & \multiline{Dual\\QAM} & \multiline{Dual\\PSK} & \multiline{Single\\QAM} & \multiline{Single\\PSK} & LAM \\\hline
		$2\times 128$ & $0.0491$ & $0.1562$ & $0.0347$ & $0.1534$ & $0.0245$ & $0.5443$\\\hline
		$128\times 2$ & $0.1627$ & $0.1562$ & $0.0347$ & $0.1534$ & $0.0245$ & $0.5443$\\\hline
		$4\times 64$ & $0.0981$ & $0.2182$ & $0.0694$ & $0.1534$ & $0.0245$ & $0.5443$\\\hline
		$64\times 4$ & $0.2615$ & $0.2182$ & $0.0694$ & $0.1534$ & $0.0245$ & $0.5443$\\\hline
		$8\times 32$ & $0.1960$ & $0.3162$ & $0.1386$ & $0.1534$ & $0.0245$ & $0.5443$\\\hline
		$32\times 8$ & $0.3318$ & $0.3162$ & $0.1386$ & $0.1534$ & $0.0245$ & $0.5443$\\\hline
		$16\times 16$ & $0.3902$ & $\mathbf{0.4472}$ & $0.2759$ & $0.1534$ & $0.0245$ & $0.5443$\\\hline
	\end{tabular}
	\label{tab:mindist8}
\end{table}

\begin{table}[!ht]
	\renewcommand{\arraystretch}{1.2}
	\parbox[t][][t]{.45\linewidth}{
		\centering
		\caption{Packing for $L=2$}
		\begin{tabular}{c|c|c}
			Bit Mapping & $\phi$ & $\vartheta$ \\\hline
			0 & $0$ & $0$ \\\hline
			1 & $0$ & $\pi$ \\\hline
		\end{tabular}
		\label{tab:3dsym2}
	}
	\hfill
	\parbox[t][][t]{.45\linewidth}{
		\centering
		\caption{Packing for $L=4$. $\alpha=\arccos\left(-\frac{1}{3}\right)$}
		\begin{tabular}{c|c|c}
			Bit Mapping & $\phi$ & $\vartheta$ \\\hline
			00 & $\frac{\pi}{2}$ & $0$ \\\hline
			01 & $0$ & $\alpha$ \\\hline
			10 & $\frac{2\pi}{3}$ & $\alpha$ \\\hline
			11 & $\frac{4\pi}{3}$ & $\alpha$ \\\hline
		\end{tabular}
		\label{tab:3dsym4}
	}
\end{table}
\begin{table}[!ht]
	\renewcommand{\arraystretch}{1.2}
	\parbox[t][][t]{.45\linewidth}{
		\centering
		\caption{Packing for $L=8$}
		\begin{tabular}{c|c|c}
			Bit Mapping & $\phi$ & $\vartheta$ \\\hline
			000 & $0$ & $\frac{\pi}{3}$ \\\hline
			001 & $\frac{\pi}{2}$ & $\frac{\pi}{3}$ \\\hline
			010 & $\frac{3\pi}{2}$ & $\frac{\pi}{3}$ \\\hline
			011 & $\pi$ & $\frac{\pi}{3}$ \\\hline
			100 & $\frac{\pi}{4}$ & $\frac{2\pi}{3}$ \\\hline
			101 & $\frac{3\pi}{4}$ & $\frac{2\pi}{3}$ \\\hline
			110 & $\frac{7\pi}{4}$ & $\frac{2\pi}{3}$ \\\hline
			111 & $\frac{5\pi}{4}$ & $\frac{2\pi}{3}$ \\\hline
		\end{tabular}
		\label{tab:3dsym8}
	}
	\hfill
	\parbox[t][][t]{.45\linewidth}{
		\centering
		\caption{Packing for $L=16$. $\alpha=\frac{2}{3}$}
		\begin{tabular}{c|c|c}
			Bit Mapping & $\phi$ & $\vartheta$ \\\hline
			0000 & $\frac{\pi}{4}$ & $\alpha$ \\\hline
			0001 & $\frac{3\pi}{4}$ & $\alpha$ \\\hline
			0010 & $\frac{7\pi}{4}$ & $\alpha$ \\\hline
			0011 & $\frac{5\pi}{4}$ & $\alpha$ \\\hline
			0100 & $0$ & $2\alpha$ \\\hline
			0101 & $\frac{\pi}{2}$ & $2\alpha$ \\\hline
			0110 & $\frac{3\pi}{2}$ & $2\alpha$ \\\hline
			0111 & $\pi$ & $2\alpha$ \\\hline
			1000 & $0$ & $\pi-\alpha$ \\\hline
			1001 & $\frac{\pi}{2}$ & $\pi-\alpha$ \\\hline
			1010 & $\frac{3\pi}{2}$ & $\pi-\alpha$ \\\hline
			1011 & $\pi$ & $\pi-\alpha$ \\\hline
			1100 & $\frac{\pi}{4}$ & $\pi-2\alpha$ \\\hline
			1101 & $\frac{3\pi}{4}$ & $\pi-2\alpha$ \\\hline
			1110 & $\frac{7\pi}{4}$ & $\pi-2\alpha$ \\\hline
			1111 & $\frac{5\pi}{4}$ & $\pi-2\alpha$ \\\hline
		\end{tabular}
		\label{tab:3dsym16}
	}
\end{table}


\end{document}